\documentclass[draft]{agujournal2019}
\usepackage{url}
\usepackage{lineno}
\usepackage[inline]{trackchanges}
\usepackage{soul}

\draftfalse

\journalname{Earth and Space Science}

\usepackage{amsmath}
\usepackage{bm}
\usepackage{multirow}
\usepackage{booktabs}
\usepackage{xcolor}
\usepackage{tikz}
\usepackage{array}
\usepackage{makecell}
\usetikzlibrary{shapes.geometric, arrows.meta, positioning, fit, backgrounds, calc}
\usepackage{ragged2e}
\usepackage{graphicx}

\begin{document}
\justifying

\title{Optimizing GEDI Simulator Configuration for European Temperate Forests}


\authors{Selim Behloul\affil{1,2,3},
Nikola Besic\affil{2,3},
Steven Hancock\affil{4,5},
Cedric Vega\affil{2,3},
Sylvie Durrieu\affil{6},
Jean-Pierre Renaud\affil{2,3,7},
Ibrahim Fayad\affil{1},
and Philippe Ciais\affil{1}}

\affiliation{1}{LSCE/IPSL, CEA-CNRS-UVSQ, Universit\'e Paris Saclay, 91191 Gif-sur-Yvette, France}
\affiliation{2}{Universit\'e de Lorraine, G\'eodata Paris, IGN, LIF, 54000 Nancy, France}
\affiliation{3}{Universit\'e Gustave Eiffel, G\'eodata Paris, IGN, LIF, 54000 Nancy, France}
\affiliation{4}{School of GeoSciences, University of Edinburgh, Edinburgh EH9 3FF, UK}
\affiliation{5}{National Centre for Earth Observation, UK}
\affiliation{6}{UMR TETIS, INRAE, AgroParisTech, CIRAD, CNRS, Univ Montpellier, F-34196, France}
\affiliation{7}{Office National des For\^ets RDI, 54600 Villers-l\`es-Nancy, France}

\correspondingauthor{Selim Behloul}{selim.behloul@lsce.ipsl.fr}


\begin{keypoints}
\item Weighting ALS returns by their recorded intensity improves GEDI waveform simulation accuracy across all seasonal conditions.
\item GEDI L2A algorithm a3 provides the best ground reference and minimizes vertical RH bias.
\item Low-sensitivity and leaf-off GEDI shots are simulated reliably, allowing their use to increase calibration and inference dataset size.
\end{keypoints}


\begin{abstract}

Accurate estimation of aboveground biomass density is essential for quantifying forest carbon stocks. NASA's GEDI mission provides valuable canopy structure data, but its sparse sampling necessitates the use of simulators to calibrate biomass models at field inventory locations. The widely used simulator of Hancock et al. (2019) emulates GEDI waveforms from airborne LiDAR point clouds, yet it has never been validated over European temperate forests. Here, we compare approximately 8,100 pairs of observed and simulated GEDI relative height (RH) profiles across French forests using the national airborne LiDAR program as input. We separate two sources of error: waveform modeling differences, assessed by referencing both simulated and real RH metrics to a common ALS-derived ground elevation, and ground detection bias, evaluated by comparing each GEDI L2A processing algorithm against the ALS reference. Under the baseline configuration, the mean absolute bias across the full RH profile reaches 0.67\,m in leaf-on and 1.20\,m in leaf-off conditions. Switching to intensity-based return weighting and selecting the a3 L2A algorithm reduces these biases to  0.43\,m and  0.42\,m respectively. The a3 algorithm also achieves near-unbiased ground detection (+0.04\,m versus -0.82\,m for the default), directly reducing a previously overlooked source of error. We also show that leaf-off acquisitions and low-sensitivity shots, both typically excluded from standard biomass products, are simulated as reliably as their counterparts, substantially expanding the potential calibration and inference datasets.

\end{abstract}

\section*{Plain Language Summary}
NASA's GEDI spaceborne laser instrument measures forest height from the International Space Station, but its footprints are sparsely distributed and cannot cover every forest inventory plot. Researchers therefore use simulators that reproduce GEDI measurements from detailed airborne laser surveys, then use these synthetic measurements to build biomass models. This study tests whether the most widely used simulator produces reliable results over French forests. Using the French national airborne laser survey as input, we generated about 8,100 synthetic GEDI measurements and compared them directly with real GEDI observations. We found that the standard simulator settings overestimate canopy  metrics because they treat every laser return equally, regardless of how much light it actually reflected. Switching to a weighting approach based on reflected intensity reduces the error substantially and removes a systematic seasonal bias that made winter (leaf-off) simulations less reliable than summer ones. With the improved settings, winter airborne surveys and low-power GEDI beams can be used alongside standard data to calibrate forest carbon models, significantly expanding the usable dataset for large-scale forest monitoring.


\section{Introduction}

Accurate estimation of aboveground biomass density (AGBD) is essential for quantifying forest carbon stocks \cite{DUNCANSON2022112845} and understanding ecosystem dynamics. Forests play a key role in the global carbon cycle \cite{ipcc2023}, and small errors in biomass estimation can propagate to large uncertainties in carbon accounting at regional or global scales. While field inventory plots provide high-accuracy measurements, their spatial and temporal coverage is limited, motivating the use of remote sensing data \cite{ARAZA2022112917} to extrapolate biomass estimates over broader areas.

NASA's Global Ecosystem Dynamics Investigation (GEDI) sensor \cite{GEDIPAPER_DUBAYAH2020100002}, onboard the International Space Station (ISS), provides canopy structure data \cite{verticalmetrics} at a global scale, enabling advances in biomass estimation and forest ecology. This full-waveform instrument records the entire backscattered LiDAR signal, and the distribution of energy provides information on the vertical structure of vegetation. From these waveforms, the vertical distribution of intercepted surfaces like canopy layers, understory and ground can be retrieved. The waveform processing first determines the useful signal extent by applying noise thresholds that exclude low-energy regions. The cumulative energy profile is then computed between these bounds. Relative height (RH) metrics are defined as the heights at which specific percentiles (e.g., 25\%, 50\%, 98\%) of the cumulative energy are reached, expressed relative to the detected ground elevation. This processing chain transforms raw waveforms (L1B product) into a RH profile (L2A product).

These RH can be converted into aboveground biomass estimates using empirical models. One of the most robust approaches remains their calibration against field inventory plots \cite{local_calibration}. GEDI operates two distinct beam types: full-power beams, which provide a higher signal-to-noise ratio (SNR), and coverage beams, which were designed to increase spatial sampling density but operate at a lower SNR. Despite this dual-beam strategy, GEDI's sampling remains sparse in both space and time: footprints are spaced approximately 60 m apart along-track and 600 m across-track, and the revisit frequency over any given inventory plot location is limited. This represents a major constraint for local calibration and validation of biomass estimation models.

Methods have been developed to link spaceborne LiDAR data with field measurements, for example by predicting GEDI RH over NFI plots to support biomass modeling and account for measurement uncertainty \cite{MAY2024103797}. Some approaches use wall-to-wall remote sensing products, such as Sentinel-1 radar and Sentinel-2 optical imagery, to link GEDI observations with field data \cite{BESIC_2024, SCHLEICH2025122964}. The forest and remote sensing community is also increasingly using simulators to generate synthetic GEDI waveforms in areas without real GEDI observations. Among the available tools, the simulator developed by \citeA{Hancock} has been widely adopted. It requires only a LiDAR point cloud as input to generate a continuous GEDI-like waveform. This tool is assumed to generate unbiased GEDI-like metrics, which are used directly for mapping vertical forest structure \cite{LAI_structure_from_simu} or to calibrate biomass models at global \cite{L4A_paper} or local scale \cite{brazil_simunfi_biomass,PASCUAL2025124313}.

However, this simulator was initially validated in a limited set of biomes, including Gabonese tropical forests and North American temperate forests, raising questions about its applicability to other forest types and environmental conditions. Only a few studies \cite{DART_GEDI} have directly compared simulated waveforms and metrics to real GEDI observations as a validation assessment rather than treating simulations as an unbiased reference. For instance, \citeA{simu_plusprecis_quereel} compared simulations with real GEDI data in Australian sclerophyll forests, finding that the simulator produced more accurate RH95 estimates than real GEDI observations, but their analysis was restricted to RH95 and RH98 and did not account for GEDI geolocation uncertainty or separate ground detection errors from waveform modeling differences. This gap in the literature highlights the need for a more comprehensive evaluation of the simulator across diverse ecosystems and input data characteristics, as any model calibrated using this simulator may inherit its potential biases. Systematic differences could then propagate into downstream products, leading to errors in biomass and carbon assessments at large scales.

The recent French national airborne LiDAR high-density (ALS) program (2020--2026) offers a complete coverage of French forests during GEDI's operational period (2018--present). In this study, we compare simulated and observed RH profiles across a wide range of forest types and phenological conditions in France using national airborne LiDAR data. Our ALS dataset was mainly acquired with a RIEGL instrument (1064\,nm) at a much higher point density than in the original simulator validation \cite{Hancock}, and we explore the influence of such dense inputs through alternative point cloud weighting strategies within the simulation. Although differences in ecological context and point density may prevent a direct comparison with the original validation, this study constitutes an independent assessment of the simulator's performance over French forests, and to our knowledge the first over European temperate forests. Unlike evergreen ecosystems, these forests undergo marked seasonal changes, meaning that both ALS acquisitions and GEDI observations may be conducted under leaf-off conditions, data that are typically excluded from studies using GEDI \cite{L4A_paper} due to concerns about their representativeness. Here, we explicitly include leaf-off data to test whether they can reliably complement leaf-on acquisitions for biomass calibration and inference \cite{noleafoffeffect}. Our final aim is to calibrate robust local biomass models by generating unbiased synthetic GEDI-like metrics at national forest inventory plot locations, and then to apply them to real GEDI shots for wall-to-wall mapping \cite{FORMS_H} or through a gridded statistical framework at kilometric resolution \cite{GEDIPAPER_DUBAYAH2020100002, Japon}. 

However, sample size remains a limiting factor for large-area biomass mapping \cite{sample_size_importance, stahl2016use, Patterson_2019}. Therefore, as a secondary objective, we take advantage of the controlled nature of simulations to directly test the widely held assumption that coverage beams under-perform relative to full-power beams \cite{GeoGEDI, Jia2025, Tong2024}. This addresses the practical question of the lowest sensitivity at which a model calibrated on simulated data can be reliably applied to real GEDI observations. Sensitivity metric quantifies the ability of a shot to detect the ground return beneath a closed canopy \cite{gedi_l2_userguide}. A value of 0.96 indicates that the ground return is detectable even under a canopy cover of up to 96\%. Using sensitivity thresholds of 0.96–0.98 \cite{Validation_africa} results in the exclusion of most coverage beams shots, which typically have lower sensitivity values than full-power beams because of their lower SNR. We therefore intentionally lower the sensitivity threshold to 0.9 to evaluate whether lower-sensitivity shots, predominantly associated with coverage beams, can be simulated as faithfully as high-sensitivity shots. If this proves to be the case, the inclusion of coverage beams could substantially expand the usable GEDI archive and provide valuable insights for the design of future spaceborne LiDAR missions.

Because simulated metrics are tied to an ALS ground and real GEDI metrics to a waveform-detected ground, systematic offsets can arise when models calibrated on simulated metrics are applied. In this study, we explicitly separate ground-referencing differences from waveform modeling errors, and identify the simulation strategy and GEDI processing algorithm that minimize both. Since simulated waveforms are noise-free, some differences from real observations are expected; our objective is to identify and understand their sources, and to provide practical recommendations for users of both the simulator and NASA GEDI datasets.

In summary, this study addresses these key questions:
\begin{enumerate}
    \item To what extent do simulated GEDI RH metrics from the simulator baseline configuration differ from real GEDI observations, and can alternative simulation configurations or L2A processing algorithms reduce these differences?
    \item What are the main drivers of these differences --- including forest type, seasonality and sensitivity?
    \item Could leaf-off acquisitions and low sensitivity shots extend the GEDI dataset used for biomass calibration and estimation?
\end{enumerate}

The \citeA{Hancock} simulator will be denoted throughout the manuscript as ``the simulator''. GEDI processing algorithms are referred to as a1--a6 in text, tables and figures to reflect the field names in the GEDI L2A product.

\clearpage

\section{Materials and Methods}

\subsection{Study Area}

Our study focuses on metropolitan France, which includes a wide range of forest types \cite{ign_memento_2025}, from temperate broadleaf forests in the north to Mediterranean woodlands in the south and a variety of topographic and climatic gradients.

\subsection{Datasets}

\textit{All GEDI products were downloaded from Earthdata Search - NASA.}

\textbf{GEDI Level 2A (L2A):} The V2 L2A products provide relative height (RH) metrics \cite{gedi_l2_userguide}, which summarize vertical canopy structure based on the cumulative energy of denoised waveforms. These RH metrics serve as the reference GEDI data, against which simulated metrics generated by the Hancock simulator are compared. Metrics are computed using specific noise thresholds that define the portion of the waveform included in the cumulative energy calculation. For each footprint, the L2A product provides six RH profiles corresponding to algorithms a1--a6 (Table \ref{tab:threshold_configs}). For convenience, the GEDI processing pipeline also designates one of these six profiles as the default per footprint, based on laser return
energy, its Plant Functional Type and regional location \cite{gedi_l2_userguide}. In our evaluation, simulated metrics are compared against the results of the six algorithms (Table \ref{tab:threshold_configs}), including the default profile, in order to comprehensively assess performance across the range of L2A processing variations. 

Auxiliary information (\textit{quality\_flag}, \textit{degrade\_flag}, \textit{sensitivity}) provided with the L2A products was also used to guide the selection of corresponding L1B waveforms (see following subsection), ensuring that only high-quality shots were included in the comparison with simulations. Beyond these quality filters, we used the \textit{leaf\_off\_flag} to stratify our analyses. The \textit{leaf\_off\_flag} is derived from the VIIRS Global Land Surface Phenology Product and indicates whether an observation was recorded during leaf-off conditions. This per-shot phenological flag enables a more accurate separation of leaf-on and leaf-off conditions than a simple assignment based on acquisition date windows, and was therefore adopted for the seasonal stratification presented in Section~\ref{sec:phenology}.

\begin{table}[htbp]
\caption{Smoothing and Threshold Settings Used in GEDI L2A Processing Algorithms (a1--a6)}
\centering
\begin{tabular}{lcccc}
\toprule
Algorithm & Smooth width & Smooth width (zcross) & Front threshold & Back threshold \\
\midrule
a1 & 6.5 & 6.5 & 3$\sigma$ & 6$\sigma$ \\
a2 & 6.5 & 3.5 & 3$\sigma$ & 3$\sigma$ \\
a3 & 6.5 & 3.5 & 3$\sigma$ & 6$\sigma$ \\
a4 & 6.5 & 6.5 & 6$\sigma$ & 6$\sigma$ \\
a5 & 6.5 & 3.5 & 3$\sigma$ & 2$\sigma$ \\
a6 & 6.5 & 3.5 & 3$\sigma$ & 4$\sigma$ \\
\bottomrule
\end{tabular}
\par\smallskip
\begin{minipage}{\linewidth}
\footnotesize \textbf{Note.} The workflow proceeds as follows: the raw waveform is first smoothed (\textit{Smooth width}) to reduce noise and help identify noise-only regions. \textit{Front} and \textit{Back thresholds}, expressed in multiples of the corrected noise standard deviation ($\sigma$), define the upper (\textit{toploc}) and lower (\textit{botloc}) bounds of the usable signal. Within this detected signal extent, the waveform is denoised using (\textit{Smooth width (zcross)}) to refine canopy and ground detection. These settings determine the portion of the waveform used for RH metric computation.
\end{minipage}
\label{tab:threshold_configs}
\end{table}

\textbf{GEDI Level 1B (L1B):} The latest V2 L1B products provide the raw waveforms necessary to collocate GEDI waveforms with their corresponding simulated GEDI signals. This step, required due to geolocation inaccuracies in GEDI footprint locations, is carried out prior to RH profile evaluation. Since L1B data do not include quality indicators, we first used L2A products to identify high-quality tracks. Only L1B tracks with at least 60\% of footprints corresponding to high-quality L2A shots (\textit{quality\_flag} = 1 and \textit{degrade\_flag} = 0) within the LiDAR-acquired area were retained. A \textit{quality\_flag} value of 1 indicates that the laser shot satisfies several quality criteria, including a \textit{sensitivity} greater than 0.9 \cite{gedi_l2_userguide}. We did not apply any additional sensitivity filtering using more restrictive thresholds commonly adopted in the literature (e.g., 0.96-0.98), in order to retain a sufficient number of coverage beam footprints and assess their performance under operational conditions.

\textbf{National LiDAR HD dataset:} We used an ongoing high-density airborne laser scanning (ALS) campaign program, aiming at covering the entire French territory with a minimum point density of 10 points per m$^2$, hereafter referred to as the French national ALS program. These LiDAR point clouds serve as the primary input for the simulator and provide a high-resolution reference for comparing both simulated and real GEDI metrics.  The dataset was acquired using multiple sensors, predominantly the RIEGL VQ-480 and the Galaxy T2000, both operating at 1064\,nm. Ground and vegetation returns were separated using the point classifications provided in the delivered LAZ files, without additional filtering or reclassification. Ground points correspond to class 2 and vegetation points to classes 3--5 (low, medium, and high vegetation respectively), following the standard LAS 1.4 classification scheme \cite{ASPRS2019}. These classifications are produced by IGN through a combination of automatic algorithms and, depending on the acquisition block, manual corrections by certified subcontractors \cite{IGN2026}. The ground classification was used to derive the ALS reference ground elevation within each simulated footprint, while the simulator itself takes the full point cloud as input.

\begin{figure}[htbp]
\noindent\includegraphics[width=\textwidth]{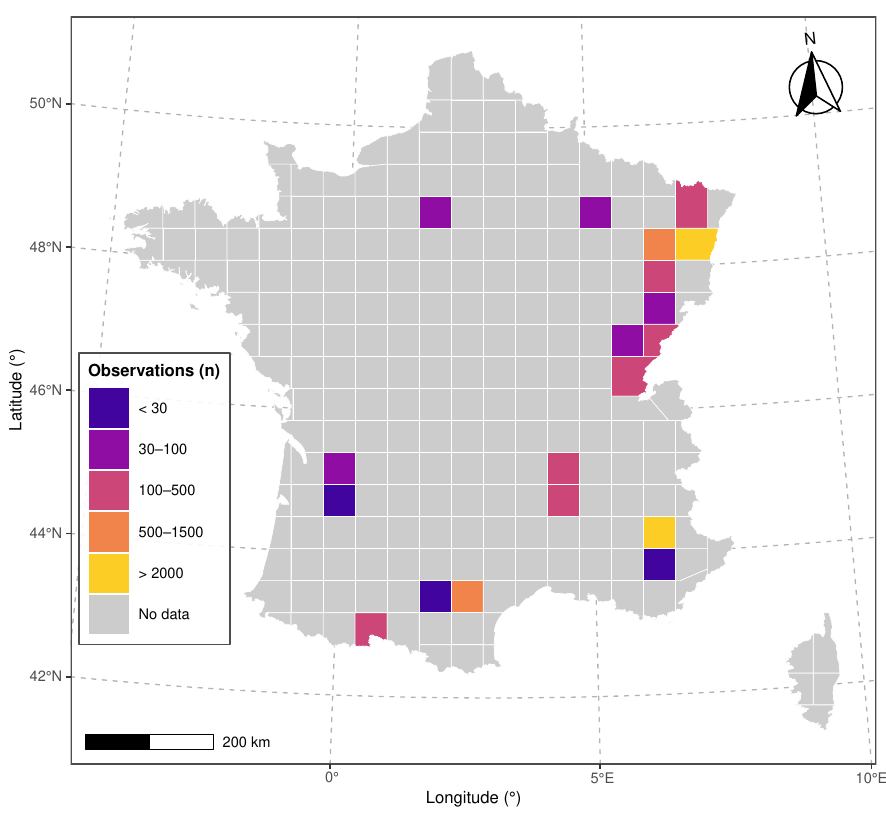}
\caption{Spatial distribution of the pairs of simulated and real waveforms across LiDAR nomenclature units.}
\label{fig:maplidarhd}
\end{figure}

Each 50$\times$50\,km unit (see Figure~\ref{fig:maplidarhd}) was acquired over a variable time window ranging from a few weeks to approximately four months, during which conditions could be either leaf-on or leaf-off depending on the season and geographic location. To minimize phenological mismatch, only GEDI shots whose acquisition date fell within the start and end dates of the corresponding LiDAR unit were retained. The temporal correspondence between ALS and GEDI acquisitions represents an inherent limitation of the French national ALS program, where tile acquisition windows can span several months. However, the original validation of the simulator imposed no temporal constraint between ALS and reference waveform data \cite{Hancock}, and simulator accuracy was shown to be robust to variations in ALS instrument configuration \cite{Hancock}. Our filtering approach therefore represents a stricter temporal control than previously applied. For the seasonal stratification, we did not rely solely on these acquisition windows; we also used the per-shot \textit{leaf\_off\_flag} from the L2A product to guide the phenological assignment of each footprint. This ensures that the seasonal separation reflects the actual phenological state at the time of each GEDI observation, rather than relying on acquisition date windows alone.

\textbf{Ancillary data:} To characterize forest type, we used the most recent Corine Land Cover dataset \cite{clc2018_raster}. This European land cover inventory has a 100\,m spatial resolution and a reference year of 2018 (revised in 2020), with an overall thematic accuracy of 92.4\% based on an independent validation across 39 European countries \cite{EEA2020validation}. For France specifically, thematic accuracy is high for broadleaf and coniferous forests ($>$\,90\%), but mixed forests are substantially under-detected, with only 69\% of actual mixed forest areas correctly identified as such, reflecting the known difficulty of delineating mixed forest polygons \cite{EEA2020validation}. Each footprint was assigned to one of three broad categories (broadleaf, coniferous, mixed). This straightforward classification was preferred to a more detailed one to maintain a balanced dataset ($\sim$3000 shots per forest type) and to ensure the potential transferability of our results to other European contexts. This stratification was essential for isolating the phenological effects.

\subsection{Indicators Derived From ALS Point Cloud}

These indicators are extracted using the \textbf{\textit{gediMetric}} tool developed within the simulator framework.
\begin{enumerate}
\item \textbf{True ground elevation:} The ground elevation was calculated as the center of gravity of all ALS ground-classified points within the simulated GEDI footprint. \\

\item \textbf{ALS canopy cover:} Canopy cover was derived from the ALS point cloud by comparing the number of canopy returns against the total number of returns (ground + canopy) within the simulated GEDI footprint. A correction factor ($\rho_{veg} / \rho_{ground} = 0.57 / 0.40$) was applied, where $\rho_{veg}$ and $\rho_{ground}$ represent the reflectance of vegetation and bare ground at the laser wavelength \cite{ARMSTON201324}, respectively, to account for the difference in return probability between these surfaces.

\end{enumerate}

\subsection{Indicator Derived From Simulated Waveform}

\begin{enumerate}

\item \textbf{Ground slope:} The effective ground slope was estimated from the width of the simulated (noise-free) ground return. It was calculated as $\theta = \arctan\left(\sqrt{\sigma_g^2 - \sigma_p^2} / \sigma_f\right)$, where $\sigma_g$ is the standard deviation of the ground peak in the waveform, $\sigma_p$ is the system pulse width, and $\sigma_f$ is the footprint width. This metric represents the combined effect of topographic slope and surface roughness within the footprint and is expressed in degrees.
\end{enumerate}

\subsection{GEDI Simulation Framework}

\subsubsection{Theoretical Basis for Waveform Simulation}

The simulator models each virtual GEDI footprint as a combination of spatial weighting, return-specific weighting, optional sampling-density normalization, vertical convolution, and instrument sampling.

\textbf{1. Spatial Weighting.}
The first step assigns each ALS point a weight based on its position relative to the center of the footprint. Points closer to the center contribute more strongly, following a Gaussian footprint model.

For a virtual footprint centered at $(x_0, y_0)$, the spatial weight of ALS point $i$ at location $(x_i, y_i, z_i)$ is:

\begin{linenomath*}
\begin{equation}
    w_i = \frac{1}{\sigma_f \sqrt{2\pi}} \,
    e^{-\frac{(x_i - x_0)^2 + (y_i - y_0)^2}{2\sigma_f^2}},
    \label{eq:footprint}
\end{equation}
\end{linenomath*}

\noindent where $\sigma_f$ is the spatial standard deviation of the Gaussian footprint, and the footprint diameter at $1/e^2$ intensity equals $4\sigma_f$.

\textbf{2. Return-specific weighting.}
Each ALS return is then assigned a factor $I_i$ describing how strongly it contributes to the simulated waveform. Three options are implemented:

\begin{linenomath*}
\[
I_i =
\begin{cases}
1, & \text{count method},\\[4pt]
1/n_{\mathrm{hits}}, & \text{frac method},\\[4pt]
\mathrm{intensity}_i, & \text{int method}.
\end{cases}
\]
\end{linenomath*}

The \texttt{count} method gives all returns equal importance; the \texttt{frac} method down-weights returns when several echoes originate from the same laser shot (each return receives $1/n_{\mathrm{hits}}$ of the pulse energy); and the \texttt{int} method weights returns proportionally to their recorded ALS intensity. All simulations were generated using the three available return-specific options (count, frac and int). Following \citeA{Hancock}, the \texttt{count} method serves as our baseline configuration.

\textbf{3. Density normalization.}
An optional normalization can be applied to correct for horizontal sampling heterogeneity within a footprint, for instance due to scan-angle attenuation, flight-line overlap, or edge-of-tile under-sampling. The simulator constructs a 2D grid over the footprint at a resolution of 1.5\,m and counts the number of ALS pulses falling within each horizontal cell (i.e., within the corresponding vertical column). Each return is then assigned:

\begin{linenomath*}
\[
D_i = \frac{1}{\rho(x_i,y_i)},
\]
\end{linenomath*}

\noindent where $\rho(x_i,y_i)$ is the local ALS pulse density. Normalizing by pulse rather than point density ensures that the relative contribution of returns is not reduced simply because multiple returns originate from the same pulse within a dense vegetation column. In this study, the density normalization option was enabled for all simulations. The final contribution of ALS return $i$ becomes:

\begin{linenomath*}
\[
I_{w,i} = w_i \, I_i \, D_i.
\]
\end{linenomath*}

\textbf{4. Vertical binning.}
The weighted contributions of all ALS returns inside the footprint are aggregated into a vertical histogram at the instrument's sampling resolution (e.g., 15\,cm for GEDI). All returns in the same bin are summed, producing a discrete profile $H[k]$.

\textbf{5. Vertical convolution with the system pulse.}
\textit{Convolution may be performed before or after discretization; the default configuration (used in this study) applies convolution on the discretized waveform, which is computationally more efficient.}

The profile is convolved with the system pulse $p(z)$ to obtain the simulated waveform:

\begin{linenomath*}
\[
I(z) = \sum_k H[k] \, p(z - z_k),
\]
\end{linenomath*}

where $z_k$ is the elevation of bin $k$. This produces a noise-free GEDI-like waveform.

\textbf{Note:} In this study, we do not use the default Gaussian pulse. Instead, all simulations rely on the empirically measured GEDI instrument pulse, which captures pulse asymmetry and detector response effects more accurately. Using the real pulse ensures that differences between simulated and observed waveforms do not originate from an idealized pulse model.

\textbf{In summary:}
\begin{enumerate}
    \item ALS points within the footprint are identified.
    \item Each point receives a spatial weight $w_i$ and a return-specific factor $I_i$.
    \item The weighted points are first binned at GEDI's vertical sampling resolution.
    \item The discretized vertical profile is then convolved with the instrument pulse shape.
\end{enumerate}

\textbf{Underlying assumptions.}
The simulator implements the large-footprint waveform modeling framework of \citeA{Blair} and relies on the following assumptions:
\begin{enumerate}
    \item the ALS point cloud provides a statistically representative sampling of non-occluded scattering surfaces at the scale of a GEDI footprint;
    \item ALS returns represent either equal intercepted areas (\texttt{count} and \texttt{frac} methods) or areas proportional to their recorded intensity (\texttt{int});
    \item energy spreading is governed by the Gaussian footprint and the instrument pulse shape, here using the empirical GEDI pulse;
    \item multiple scattering and reflectance anisotropy are ignored.
\end{enumerate}

\subsubsection{Simulation Tools and Processing Workflow}
\label{sec:simulation_tools}

The GEDI simulation framework provides a set of tools to generate, align, and analyze simulated GEDI-like waveforms from ALS data. The two specific tools used in this study are \textbf{\textit{collocateWaves}} and \textbf{\textit{gediMetric}}.

\begin{enumerate}
    \item \textbf{\textit{collocateWaves}}. Real GEDI footprints are subject to horizontal geolocation uncertainties on the order of 10~m \cite{gedi_l2_userguide}. Using the raw footprint locations would introduce systematic misalignment. This tool corrects geolocation uncertainties by exploring a user-defined horizontal and vertical search grid and generating simulated waveforms at each candidate offset location. For an entire GEDI L1B track, the algorithm identifies the optimal displacement $(\Delta x, \Delta y, \Delta z)$ that maximizes the overall Pearson correlation between real GEDI waveforms and those simulated from ALS data. This produces a consistent spatial alignment between simulated and observed data. Following recommendations from previous studies \cite{GeoGEDI}, we adopt a $20\,\text{m} \times 20\,\text{m}$ horizontal search window. Note that this geolocation correction is based on correlation between real and simulated waveforms using the \texttt{count} weighting. Other weighting methods are derived afterwards using the same optimal displacement. \\

    In practice, the collocation step is sensitive to noise, particularly when estimating the vertical offset ($\Delta z$) between the WGS84 ellipsoid used by GEDI and the orthometric reference frame of the ALS data (IGN69 in our case). Poor-quality or degraded waveforms can cause the optimization to perform poorly or fail to converge. To mitigate this, we retain only high-quality GEDI tracks, specifically those for which the L2A product indicates \texttt{quality\_flag = 1}, \texttt{degrade\_flag = 0} for at least 60\% of the shots within the LiDAR-acquired area. The L2A product therefore guides our selection of which GEDI L1B tracks can be reliably collocated. \\

    The number of footprints collocated simultaneously depends on two interrelated factors: the number of valid L1B shots available within the input ALS tiles, and the number of ALS tiles that can be loaded at once given the considerable RAM requirements of the procedure. Under these constraints, collocation is performed on sub-tracks rather than entire L1B tracks at once, each comprising the footprints available within a given set of ALS tiles. Sub-tracks typically contain around 30 footprints. In 25 cases out of 353, fewer than 20 footprints were available due to an insufficient number of valid L1B waveforms within the ALS tiles, corresponding to 149 shots in total. \\

    The success of each alignment is assessed through the optimal correlation of the sub-track: sub-tracks with a Pearson correlation lower than 70\% are considered to have failed, and all their footprints are discarded. When the optimal offset lies near the boundary of the search window, the simplex algorithm may extend the search beyond the initial \(20\,\text{m}\) limit. The simplex stops when its size falls below the convergence tolerance (\texttt{optTol} parameter, default = 0.2) or when the maximum number of iterations (\texttt{maxIter} parameter, default = 300) is reached. In our case, 7 sub-tracks required such an extension, corresponding to 190 footprints (2\% of the dataset), with maximum observed offsets of 28.2\,m in $x$ and 26.9\,m in $y$; all achieved waveform correlations above 91\%. \\

    \item \textbf{\textit{gediMetric}} --- processes simulated large-footprint LiDAR waveforms to extract GEDI-like forest structure metrics, including relative height (RH) percentiles, canopy cover, and foliage height diversity. The tool also allows sensitivity analysis through two mechanisms:
    \begin{enumerate}
        \item Adding synthetic Gaussian noise scaled to achieve a target link margin (SNR) corresponding to GEDI's expected beam sensitivity.
        \item Applying a denoising threshold so that any bins in the waveform below this value are set to zero. The threshold is defined as:
    \end{enumerate}

\begin{linenomath*}
    \begin{equation}
        \tau = \mu + \alpha \, \sigma,
    \end{equation}
\end{linenomath*}

    where $\mu$ and $\sigma$ are the mean and standard deviation of waveform bins sampled from the first and last 20 meters of the simulated waveform. For noisy simulated waveforms, these segments are assumed to contain mostly background noise. In contrast, for noise-free simulated waveforms, they consist of low-amplitude values arising from the Gaussian nature of the convolution, rather than from stochastic noise. Parameter $\alpha$ is a user-defined scaling factor (\textit{-varScale} parameter). This same threshold $\tau$ serves both to remove noise (by zeroing bins below it) and to define the vertical extent of the signal (by identifying the lowest and highest elevations with amplitude $> \tau$).
\end{enumerate}

\subsection{Workflow}

\definecolor{col_data}{RGB}{52, 90, 130}
\definecolor{col_tool}{RGB}{255,255,255}
\definecolor{col_choice}{RGB}{230,115,30}
\definecolor{col_output}{RGB}{60,130,80}
\definecolor{col_border}{RGB}{80,80,80}
\definecolor{col_bg}{RGB}{240,243,248}
\definecolor{col_arrow}{RGB}{80,80,80}

\tikzset{
  databox/.style={
    rectangle, rounded corners=4pt,
    fill=col_data, draw=col_data!70!black, line width=0.6pt,
    text=white, font=\small\bfseries,
    minimum width=3.0cm, minimum height=0.85cm, align=center
  },
  toolbox/.style={
    rectangle, rounded corners=4pt,
    fill=col_tool, draw=col_border, line width=0.8pt,
    text=col_border, font=\small\bfseries,
    minimum width=4.6cm, minimum height=0.85cm, align=center
  },
  choicebox/.style={
    rectangle, rounded corners=3pt,
    fill=col_choice!12, draw=col_choice, line width=0.7pt,
    text=col_choice!80!black, font=\footnotesize\itshape,
    minimum width=2.8cm, minimum height=0.7cm, align=center
  },
  outbox/.style={
    rectangle, rounded corners=4pt,
    fill=col_output!18, draw=col_output!70!black, line width=0.8pt,
    text=col_output!60!black, font=\small\bfseries,
    minimum width=4.6cm, minimum height=0.85cm, align=center
  },
  seclabel/.style={
    font=\footnotesize\bfseries\color{col_border!60},
    align=right, text width=1.6cm
  },
  arrow/.style={
    -{Latex[length=2.5mm, width=2mm]},
    line width=0.9pt, color=col_arrow
  },
  thinconnect/.style={
    -{Latex[length=2mm, width=1.5mm]},
    line width=0.6pt, color=col_choice!80, dashed
  },
  thinarrow/.style={
    -{Latex[length=2mm, width=1.5mm]},
    line width=0.6pt, color=col_data!60
  },
}

\begin{figure}[htbp]
\resizebox{\textwidth}{!}{%
\begin{tikzpicture}

\node[databox] (als) at (-3.8, 0) {ALS point clouds\\{\footnotesize French LiDAR HD}};
\node[databox, right=0.5cm of als] (l1b) {GEDI L1B\\{\footnotesize raw waveforms}};
\node[databox, right=0.5cm of l1b] (l2a) {GEDI L2A\\{\footnotesize RH metrics $\times$ 6 algos}};
\node[databox, right=0.5cm of l2a] (anc) {Ancillary data\\{\footnotesize Corine LC, slope, CC}};
\node[seclabel] at (-6.6, 0) {\textsc{Inputs}};

\node[toolbox] (colloc) at (0, -2.0)
  {\texttt{collocateWaves}\\{\footnotesize 3-D geolocation correction}};

\draw[arrow] (als.south)  -- ++(0,-0.28) -| ([xshift=-0.6cm]colloc.north);
\draw[arrow] (l1b.south)  -- ++(0,-0.28) -| ([xshift=+0.1cm]colloc.north);

\begin{scope}[on background layer]
  \node[fill=col_bg, rounded corners=6pt, inner sep=9pt, fit=(colloc)] (s1bg) {};
\end{scope}
\node[seclabel] at (-3.75, -2.0) {\textsc{Stage 1}\\[-1pt]Collocation};

\node[choicebox, right=1.1cm of colloc, minimum width=3.2cm,
      yshift=0cm] (weighting)
  {Outputs 3 weightings:\\[2pt]\texttt{count} $\cdot$ \texttt{frac} $\cdot$ \texttt{int}\\
   {\tiny (count used for collocation)}};
\draw[thinconnect] ([xshift=0.3cm]colloc.east) -- (weighting.west);

\node[toolbox] (gedimet) at (0, -3.8)
  {\texttt{gediMetric}\\{\footnotesize RH metric extraction}};
\node[choicebox, right=1.1cm of gedimet, minimum width=3.0cm] (alsground)
  {Ground reference:\\[2pt]ALS-derived\\{\footnotesize (isolates waveform errors)}};
\draw[thinconnect] (alsground.west) -- (gedimet.east);
\draw[arrow] (colloc.south) -- (gedimet.north);

\begin{scope}[on background layer]
  \node[fill=col_bg, rounded corners=6pt, inner sep=9pt,
        fit=(gedimet)(alsground)(weighting)] (s2bg) {};
\end{scope}
\node[seclabel] at (-3.75, -3.8) {\textsc{Stage 2} \\[-1pt]RH Computation};

\node[toolbox] (compare) at (0, -7.0)
  {RH profile comparison\\{\footnotesize simulated vs.\ observed}};

\node[choicebox, right=1.37cm of compare, minimum width=3.0cm] (algos)
  {L2A algorithms:\\[2pt]\texttt{a1 a2 a3 a4 a5 a6}\\+ default};
\draw[thinconnect] (algos.west) -- (compare.east);

\coordinate (l2adown) at ([yshift=-0.4cm]l2a.south);
\coordinate (algostop) at ([yshift=+0.3cm]algos.north);
\draw[thinarrow] (l2a.south) -- (l2adown) -- ([xshift=3cm]l2adown -| algos.north) |- (algos.east);

\node[choicebox, left=1.1cm of compare, minimum width=3.0cm] (drivers)
  {Stratified by:\\[2pt]phenology $\cdot$ forest type\\ sensitivity };
\draw[thinconnect] (drivers.east) -- (compare.west);

\coordinate (ancdown) at ([yshift=-0.4cm]anc.south);
\coordinate (driverstop) at ([yshift=+0.3cm]drivers.north);
\draw[thinarrow] (anc.south) -- (ancdown) -- (ancdown -| drivers.north) -- (driverstop) -- (drivers.north);

\draw[arrow] (gedimet.south) -- (compare.north);

\node[outbox, line width=1.5pt] (optconf) at (0, -8.6)
  {Optimal configuration};
  
  
\draw[arrow] (compare.south) -- (optconf.north);

\node[outbox, right=0.535cm of optconf] (groundres)
  {Ground detection accuracy\\{\footnotesize per L2A algorithm}};

\draw[thinconnect] (algos.south) -- (groundres.north);

\begin{scope}[on background layer]
  \node[fill=col_bg, rounded corners=6pt, inner sep=9pt,
        fit=(compare)(groundres)(algos)(drivers)] (s3bg) {};
\end{scope}
\node[seclabel] at (-3.75, -8.6) {\textsc{Stage 3}\\[-1pt]Comparison\\[-1pt]\& Drivers};

\node at (0.7, -10.5) {
  \begin{tikzpicture}[node distance=0.4cm]
    \node[databox, minimum width=2.0cm, minimum height=0.5cm,
          font=\footnotesize\bfseries] (ld) {Input data};
    \node[toolbox, minimum width=2.0cm, minimum height=0.5cm,
          font=\footnotesize\bfseries, right=0.4cm of ld] (lt) {Tool / process};
    \node[choicebox, minimum width=2.8cm, minimum height=0.5cm,
          font=\footnotesize\itshape, right=0.4cm of lt] (lc) {Methodological choice};
    \node[outbox, minimum width=2.0cm, minimum height=0.5cm,
          font=\footnotesize\bfseries, right=0.4cm of lc] (lo) {Output / result};
  \end{tikzpicture}
};
\end{tikzpicture}
}
\caption{Overview of the methodological workflow.}
\label{fig:workflow}
\end{figure}
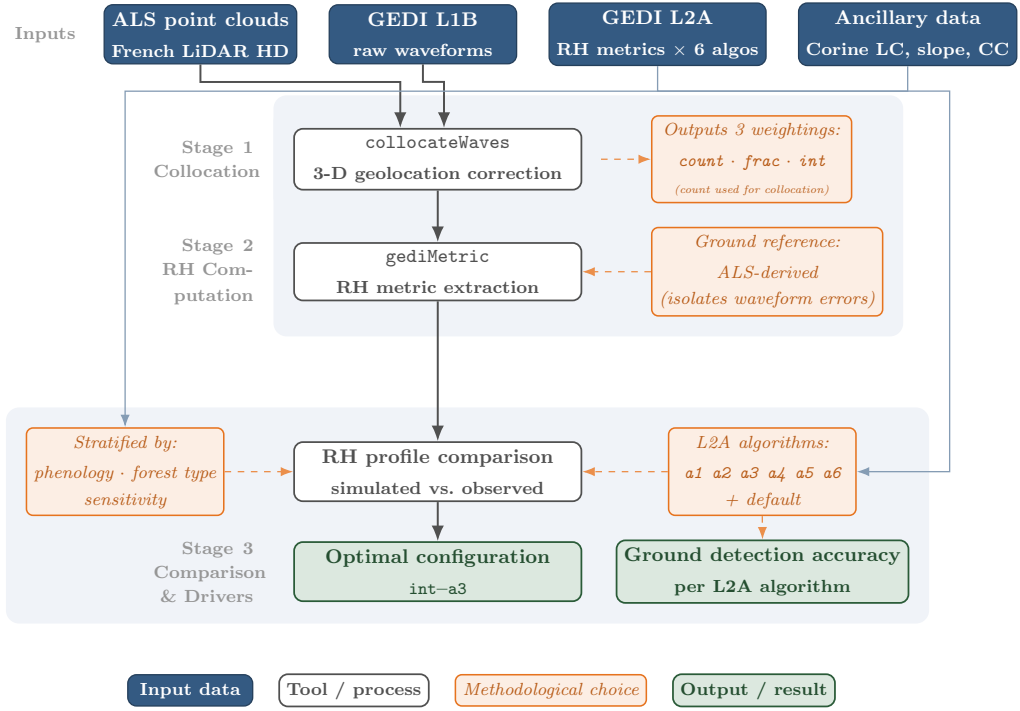

Regarding our methodology, Figure~\ref{fig:workflow} illustrates the workflow proposed in this study. We aim to separate two main sources of error: inaccuracies in waveform modeling, which affect the simulated RH metrics themselves, and systematic offsets introduced during model calibration and application. Indeed, a practical mismatch arises when simulated metrics are used to calibrate models that are then applied to real GEDI observations. Simulated RH metrics are typically referenced to an ALS-derived ground elevation, whereas real GEDI RH metrics are referenced to a ground elevation detected within the waveform itself \cite{PASCUAL2025124313, DUNCANSON2022112845}. This discrepancy can introduce systematic offsets that propagate into calibrated models. Most users rely on the default GEDI Level 2A product without exploring the six available processing algorithms. Since each algorithm applies a distinct approach to defining the useful signal extent and handling noise, they differ both in their ground elevation accuracy and their RH metric outputs.

We first evaluate each algorithm's ground elevation accuracy against ALS-derived ground to identify the most accurate. We then explore three simulation strategies, which differ in how ALS point returns are weighted and therefore influence waveform shape. By referencing all outputs, simulated and algorithm-derived, to the common ALS ground, we separate ground detection errors from waveform modeling differences and directly compare all simulation--algorithm pairings to find the closest agreement. The best-performing configuration, i.e., the one that maximizes the transferability of models calibrated on simulated metrics to real GEDI observations, is then retained and systematically compared against the baseline throughout the paper across forest types, phenological conditions, and canopy cover.

\subsubsection{Datasets Summary}

Three simulation datasets were generated using the three return-weighting schemes implemented in the simulator: \texttt{count}, \texttt{int}, and \texttt{frac}. Although we initially explored varying the \texttt{varScale} parameter to adjust the signal extent used for RH computation --- seeking a closer match with the default GEDI L2A processing --- this analysis did not yield any systematic improvement. We therefore adopt the default setting (\texttt{varScale = 3}) for all simulations. Each of the three simulated datasets was compared both against the default GEDI L2A configuration and against each of the six L2A algorithms (a1--a6) individually.

L1B and L2A footprints were also restricted to temporal overlap with LiDAR HD acquisitions to minimize phenological differences. We retained only GEDI shots whose acquisition date fell within the start and end dates of the corresponding LiDAR HD tile (50$\times$50\,km), whose acquisition window could span from a few weeks up to four months. Additional filtering removed footprints with incomplete ALS coverage (e.g., near tile edges) and retained only footprints with canopy cover strictly between 0 and 1 ($0 < \text{CC} < 1$), excluding pure ground and fully closed canopy shots. 
Finally, we retained only pairs for which the simulated RH100, referenced to the ALS-derived ground elevation, ranged between 5\,m and 60\,m. This final filtering step removes potential misclassifications in the Corine Land Cover dataset and excludes physically implausible values, as tree heights exceeding 60 m are not expected in France.

This workflow yields a homogeneous, high-confidence dataset of spatially and temporally aligned GEDI--LiDAR pairs suitable for rigorous evaluation of the simulator across diverse forest environments. Table~\ref{tab:forest_beam_summary} summarizes the dataset at the shot level, reporting for each forest and beam type the number of simulation--real data pairs, mean canopy cover (CC), mean ground slope, and the proportions of algorithms selected as the default configuration in the downloaded v2 release of GEDI products.

\begin{table}[htbp]
\caption{Final Dataset Distribution After Filtering Using Flags From the GEDI Default Configuration}
\centering
\begin{tabular}{llrrrrr}
\toprule
Forest type & Beam & N & Leaf-off (\%) & CC & Slope ($^{\circ}$) & a1 (\%) \\
\midrule
Broadleaf   & Coverage   & 1394 & 61.8 & 0.62 & 19 & 28 \\
Broadleaf   & Full power & 1487 & 57.2 & 0.63 & 19 & 44 \\
Coniferous  & Coverage   & 1481 & 31.1 & 0.64 & 15 & 30 \\
Coniferous  & Full power & 1512 & 22.2 & 0.65 & 19 & 43 \\
Mixed       & Coverage   &  781 & 54.8 & 0.64 & 21 & 22 \\
Mixed       & Full power & 1404 & 35.4 & 0.63 & 21 & 37 \\
\midrule
\textbf{Total} & & \textbf{8059} & \textbf{42.6} & \textbf{0.63} & \textbf{19} & \textbf{35} \\
\bottomrule
\end{tabular}
\par\smallskip
\begin{minipage}{\linewidth}
\footnotesize \textbf{Note.} N is the number of simulation--real GEDI pairs. Leaf-off (\%) is the proportion of pairs acquired under leaf-off conditions. CC and slope are derived from ALS data. a1 (\%) is the proportion of footprints assigned to algorithm a1 in the GEDI default configuration; the remaining footprints are assigned to a2. Dataset size may vary depending on the GEDI algorithm used and its associated quality flags.
\end{minipage}
\label{tab:forest_beam_summary}
\end{table}

\subsection{Comparison Framework and Evaluation Strategy}

\subsubsection{Signal Overall Comparison}

Waveform similarity between simulated and observed GEDI shots was evaluated through the correlation-based collocation procedure. For each subset of a GEDI track, the \texttt{collocateWaves} algorithm explored a 3D search grid and identified the displacement $(\Delta x, \Delta y, \Delta z)$ that maximized the mean Pearson correlation across all valid footprints. Mean correlations before and after collocation were compared to quantify the improvement in track-level alignment. This optimization primarily assesses agreement in vertical structure rather than absolute energy values.

\subsubsection{GEDI L2A Algorithm Evaluation}

The six GEDI L2A algorithms (a1--a6) are evaluated along two complementary axes. 

\textbf{Ground Detection accuracy.}
Once accurate spatial alignment is ensured, the ALS-derived ground elevation serves as a reference against which the ground elevation retrieved by each L2A algorithm is compared. This comparison identifies which algorithm achieves the most accurate ground detection. It addresses a practical issue: since simulated RH metrics are referenced to the ALS ground while real GEDI RH metrics are referenced to a waveform-detected ground, any algorithm-dependent ground detection bias propagates directly into all RH metrics and would introduce a systematic offset when calibrated models are applied to real GEDI observations.

\textbf{RH profile agreement.}
Beyond ground detection, each algorithm applies distinct denoising parameters and thresholds that influence the vertical energy distribution of the cumulative RH profile. We therefore examine which algorithm, combined with each of the three return-weighting strategies, yields the closest agreement between simulated and observed RH profiles. The configuration offering the best overall compromise across both axes is then retained for all subsequent analyses.

\subsubsection{RH Metrics Comparison}

\textbf{Vertical alignment.}
All RH metrics from real GEDI observations were vertically realigned to the ALS-derived ground elevation in order to remove differences caused by ground-detection errors. Since simulated waveforms are already referenced to the ALS ground elevation, only the observed GEDI RH metrics required correction. For each footprint and for each L2A algorithm, we applied a simple vertical translation equal to the difference between the ground elevation reported by that algorithm ($z_{\mathrm{ground,\,GEDI}}$) and the ALS ground elevation ($z_{\mathrm{ALS}}$):

\begin{linenomath*}
\[
\mathrm{RH}_{p}^{\mathrm{GEDI,\,aligned}}
= \mathrm{RH}_{p}^{\mathrm{GEDI}}
+ \left( z_{\mathrm{ground,\,GEDI}} - z_{\mathrm{ALS}} \right).
\]
\end{linenomath*}

This alignment ensures that comparisons isolate differences in canopy energy distribution rather than artifacts arising from algorithm-dependent ground detection.

\textbf{RH metrics of interest.}
RH0 and RH100 were excluded because their estimation is highly sensitive to noise and thresholding decisions. For visualization, most figures focus on a representative subset of RH metrics: RH98, RH90, RH70, RH50, RH25, RH10, and RH5. These seven percentiles provide a balanced summary of the vertical distribution --- three above and three below RH50 --- and correspond to the metrics most frequently used in the literature. Each figure also reports an aggregated performance metric summarizing accuracy across the full RH1--RH99 profile. It is calculated as the mean absolute bias averaged across all percentiles from RH1 to RH99.

All statistics were computed independently for each RH percentile (i.e., aggregating across footprints for a fixed height percentile). This design allows us to diagnose whether the simulator systematically over- or under-estimates specific portions of the vertical profile across diverse conditions.

\textbf{Metrics for RH profile comparison.}
Agreement between simulated and observed RH metrics was quantified using the mean bias and the root mean squared error (RMSE), which summarize systematic offset and overall error magnitude, respectively. While these metrics provide a concise measure of overall agreement, they do not distinguish between different sources of discrepancy. To gain deeper insight, we analyze the mean squared deviation (MSD), following the decomposition of \citeA{Kobayashi}, which partitions total error into three components, presented in the same order as in the equation below:

\begin{linenomath*}
\[
\text{MSD} = (\bar{y} - \bar{\hat{y}})^2 + (s_y - s_{\hat{y}})^2 + 2(1 - r)s_y s_{\hat{y}}.
\]
\end{linenomath*}

where $\bar{y}$ and $\bar{\hat{y}}$ are the mean values of observed and simulated RH metrics respectively, $s_y$ and $s_{\hat{y}}$ are their standard deviations, and $r$ is the Pearson correlation coefficient.

\begin{enumerate} 

\item \textbf{Squared Bias (SB: $(\bar{y} - \bar{\hat{y}})^2$}): systematic differences in mean RH values. \\

\item \textbf{Squared difference between standard deviations (SDSD: $(s_y - s_{\hat{y}})^2$}): mismatch in variability. \\ 

\item \textbf{Lack of correlation weighted by the standard deviations (LCS: $2(1 - r)s_y s_{\hat{y}}$}): lack of correlation between simulated and observed values.

\end{enumerate}

Together, these components provide a diagnostic framework for identifying whether discrepancies arise primarily from bias, variability mismatch, or reduced footprint-to-footprint correspondence.

\subsubsection{Stratification of the Analysis}

Once the optimal simulation configuration is identified, residual biases are examined along three stratification axes: forest type (broadleaf, coniferous, and mixed), phenological condition (leaf-on vs. leaf-off), and sensitivity group (low-sensitivity: $0.90 \le \text{sens} \le 0.96$; high-sensitivity: $0.96 < \text{sens}$). This stratification aims to identify the main drivers of residual discrepancies between simulated and observed RH profiles across diverse forest environments.

The MSD decomposition is first used to select the optimal configuration. However, because SB reflects the squared contribution of bias rather than the bias itself, it visually under-represents residual errors when their magnitude falls below 1\,m. For this reason, the stratified analyses are conducted by evaluating the bias directly in meters, which provides a more interpretable measure of simulations agreement with real GEDI measurements.

\clearpage

\section{Results}

\subsection{Collocation Results}

The correlation-based collocation procedure improved the spatial alignment between simulated and observed GEDI waveforms. Across the 32 processed GEDI tracks (approximately 10,300 shots before final quality filtering), the initial Pearson correlation between real L1B waveforms and noise-free simulations averaged 64\%. After the 3D search grid, the optimal displacement $(\Delta x, \Delta y, \Delta z)$ increased the mean correlation to 90\%, corresponding to an average improvement of +26 percentage points.

The horizontal shifts required to maximize correlation were consistent across tracks, with mean absolute offsets of 7.3~m in the east--west direction and 6.8~m in the north--south direction. These values closely match the $\sim$10~m geolocation uncertainty reported for GEDI V2 products \cite{gedi_l2_userguide}. Vertical adjustments were also stable across tracks, with $\Delta z$ values centred near -48~m, reflecting the systematic difference between the WGS84 ellipsoid used by GEDI and the IGN69 orthometric datum used in the ALS data.

We did not perform waveform-by-waveform comparisons beyond correlation-based collocation because real GEDI waveforms contain instrument noise and threshold-dependent denoising, whereas simulated waveforms are noise-free. Correlation was used solely for spatial collocation, not for quantitative evaluation. For this reason, RH metrics, which summarize the vertical energy distribution and constitute the official GEDI L2A product, provide the only physically consistent basis for evaluating the simulator. A few illustrative examples of simulated and real waveforms are nevertheless provided in ~\ref{sec:waveformsplots} to illustrate typical agreements and discrepancies at the signal level. Overall, the collocation step ensured high-quality spatial correspondence between simulated and real GEDI signals, providing a reliable foundation for subsequent comparisons.

\subsection{Ground Detection Accuracy Across GEDI Algorithms}
\label{sec:ground_detection}

The distribution of ground elevation residuals (GEDI-detected ground minus ALS-derived ground) for each L2A algorithm is illustrated in Table~\ref{tab:ground_sensitivity}. In the following, we focus in particular on the $ 0.90 \le \text{sens}$ sensitivity subset, which provides the most balanced representation of algorithm behaviour. Within this group, algorithm a5 performs most poorly, with a mean bias of -5.43\,m (RMSE\,=\,7.93\,m), due to its permissive back threshold (Table~\ref{tab:threshold_configs}), which causes the algorithm to place the ground return too low within the waveform. The default GEDI configuration and algorithm a2 yield very similar negative biases of approximately -0.82\,m (RMSE\,=\,4.07\,m) and -0.79\,m (RMSE\,=\,4.03\,m) respectively, which is expected because a2 predominates in the default configuration (Table~\ref{tab:forest_beam_summary}). Algorithms a1 and a4 produce identical ground elevation outputs, as expected from their shared back-threshold and denoising parameters (Table~\ref{tab:threshold_configs}).

\begin{table}[htbp]
\caption{Ground Detection Performance per Algorithm and Sensitivity Subset}
\centering
\begin{tabular}{l ccc ccc ccc}
\toprule
& \multicolumn{3}{c}{$0.90 \le \text{sens} \le 0.96$} & \multicolumn{3}{c}{$0.96 < \text{sens} $} & \multicolumn{3}{c}{$0.90 \le \text{sens} $} \\
\cmidrule(lr){2-4} \cmidrule(lr){5-7} \cmidrule(lr){8-10}
\textbf{Algorithm} & $n$ & \textbf{Bias} & \textbf{RMSE} & $n$ & \textbf{Bias} & \textbf{RMSE} & $n$ & \textbf{Bias} & \textbf{RMSE} \\
\midrule
\textbf{default} & 1988 & \textbf{-0.14} & \textbf{2.70} & 6071 & \textbf{-1.04} & \textbf{4.43} & 8059 & \textbf{-0.82} & \textbf{4.07} \\
a1      & 4239 & 1.70 & 5.28 & 3508 & 1.57 & 5.38 & 7747 & 1.64 & 5.32 \\
a2      & 800  & -0.87 & 3.22 & 7435 & -0.78 & 4.11 & 8235 & -0.79 & 4.03 \\
\textbf{a3}      & 3893 & \textbf{0.70} & \textbf{4.70} & 4246 & \textbf{-0.57} & \textbf{4.39} & 8139 & \textbf{0.04} & \textbf{4.54} \\
a4      & 4239 & 1.70 & 5.28 & 3508 & 1.57 & 5.38 & 7747 & 1.64 & 5.32 \\
a5      & 82   & -2.64 & 4.48 & 8141 & -5.46 & 7.96 & 8223 & -5.43 & 7.93 \\
a6      & 2358 & -0.82 & 3.67 & 5873 & -1.65 & 4.98 & 8231 & -1.41 & 4.65 \\
\bottomrule
\end{tabular}
\par\smallskip
\begin{minipage}{\linewidth}
\footnotesize \textbf{Note.} The balance between low- and high-sensitivity shots varies because each algorithm has its own sensitivity estimation. Sample sizes correspond to the \texttt{count} weighting.
\end{minipage}
\label{tab:ground_sensitivity}
\end{table}

Among all six algorithms, a3 yields the most accurate ground elevation estimates, with a near-zero mean bias of +0.04\,m and an RMSE of 4.54\,m. This result is particularly noteworthy given that the default GEDI V2 product relies exclusively on a1 and a2 across our dataset (Table~\ref{tab:forest_beam_summary}), indicating that significant gains in ground detection accuracy could be achieved simply by selecting a3 as the reference algorithm.

A more detailed inspection of the two sensitivity subsets reveals that a3 achieves a well-balanced performance across sensitivity groups, with biases of +0.70\,m and -0.57\,m for low- and high-sensitivity shots respectively, similar in magnitude but opposite in sign. This balance reflects a fundamental difference in sensitivity classification: a3 distributes difficult waveforms more evenly across both groups, whereas the default configuration concentrates them in the high-sensitivity subset (-1.04\,m), while keeping the low-sensitivity group (-0.14\,m) artificially clean.

Now that we have identified a3 as the best-performing algorithm for ground detection, we turn to the question of which algorithm best represents the observed canopy energy distribution. 

\subsection{Comparison of Simulated RH Profiles Across GEDI Algorithms}
\label{sec:simu_vs_algo}

An overview of the MSD decomposition over RH percentiles 1--99 is provided in Table~\ref{tab:msd_decomposition_full}, covering each combination of simulation method (\texttt{count}, \texttt{int} and \texttt{frac}) and L2A algorithm. Across all scenarios, the lack-of-correlation term (LCS) dominates the total error, accounting for approximately 90\,\% of the MSD, while the squared-bias (SB) and standard-deviation mismatch (SDSD) terms contribute comparatively little.

\begin{table}[htbp]
\caption{Mean MSD Decomposition (SB, SDSD, LCS, MSD) Averaged Over RH Percentiles 1--99, for Each Combination of Return-Weighting Method and L2A Processing Algorithm.}
\centering
\fontsize{9}{11}\selectfont
\begin{tabular}{l c c c c c c c}
\toprule
\textbf{Metric} & \textbf{default} & \textbf{a1} & \textbf{a2} & \textbf{a3} & \textbf{a4} & \textbf{a5} & \textbf{a6} \\
\midrule
SB count   & \textbf{0.86} & 0.41 & 1.03 & 0.45 & 1.79 & 1.50 & 0.76 \\
SB frac    & 0.59 & 0.19 & 0.72 & 0.20 & 1.28 & 1.17 & 0.47 \\
SB int     & 0.30 & 0.10 & 0.37 & \textbf{0.09} & 0.65 & 0.73 & 0.20 \\
\midrule
SDSD count & \textbf{0.10} & 0.30 & 0.10 & 0.33 & 0.34 & 0.13 & 0.13 \\
SDSD frac  & 0.12 & 0.16 & 0.12 & 0.17 & 0.18 & 0.17 & 0.11 \\
SDSD int   & 0.10 & 0.15 & 0.11 & \textbf{0.17} & 0.16 & 0.15 & 0.09 \\
\midrule
LCS count  & \textbf{8.46} & 8.75 & 8.63 & 9.00 & 12.47 & 8.78 & 8.63 \\
LCS frac   & 7.24 & 7.50 & 7.32 & 7.72 & 10.59 & 7.48 & 7.33 \\
LCS int    & 6.86 & 7.14 & 6.90 & \textbf{7.35} & 9.66 & 7.04 & 6.93 \\
\midrule
MSD count  & \textbf{9.42} & 9.46 & 9.76 & 9.78 & 14.61 & 10.41 & 9.53 \\
MSD frac   & 7.95 & 7.85 & 8.17 & 8.10 & 12.05 & 8.82 & 7.90 \\
MSD int    & 7.26 & 7.39 & 7.38 & \textbf{7.61} & 10.47 & 7.92 & 7.22 \\
\bottomrule
\end{tabular}
\par\smallskip
\begin{minipage}{\linewidth}
\footnotesize \textbf{Note.} All values in m$^2$. RH metrics are referenced to the ALS-derived ground elevation. Results are for $0.90 \le$ sensitivity. Sample sizes are reported for each algorithm using the \texttt{count} weighting: default (8,059), a1 (7,747), a2 (8,235), a3 (8,139), a4 (7,747), a5 (8,223), a6 (8,231). $N$ varies mainly across algorithms due to algorithm-specific quality and sensitivity filters,  with only minor variations across weighting schemes.
\end{minipage}
\label{tab:msd_decomposition_full}
\end{table}

Within the \texttt{count} baseline compared against the default GEDI configuration, SB reaches 0.86 and total MSD reaches 9.42. Switching to the int weighting systematically reduces both SB and MSD across all algorithms, reflecting a better match between the vertical energy distributions of simulated and observed waveforms. The combination of \texttt{int} weighting with algorithm a3 achieves the lowest squared bias (SB = 0.09) among all tested configurations, while maintaining a competitive MSD of 7.61.

Tables~\ref{tab:msd_decomposition_low_sensitivity_N_inline} and~\ref{tab:msd_decomposition_high_sensitivity_N_inline} (\ref{sec:low_high_appendix}), which report results for low- and high- sensitivity shots, respectively,  show that the \textit{int--a3} configuration performs consistently across the range of signal-to-noise conditions. For high-sensitivity shots ($>$0.96) the root-mean-square error (RMSE) of the RH profile is 2.55\,m (MSD\,=\,6.53); for low-sensitivity shots (0.9--0.96) it rises moderately to 3.0\,m (MSD\,=\,8.78). As observed for ground detection (Section~\ref{sec:ground_detection}), the advantage of a3 over the default configuration is primarily driven by high-sensitivity shots, while the default achieves a smaller bias in the low-sensitivity subset (see \textit{int--default} and \textit{int--a3}: Tables~\ref{tab:msd_decomposition_high_sensitivity_N_inline} -~\ref{tab:msd_decomposition_low_sensitivity_N_inline} in \ref{sec:low_high_appendix}). Considering all shots together ($0.90 \le \text{sens}$, Table~\ref{tab:msd_decomposition_full}), a3 provides both the most accurate ground detection (Section~\ref{sec:ground_detection}) and the lowest RH bias across all weighting strategies. We therefore select the combination of \texttt{int} weighting with a3, denoted \textit{int--a3}, as our optimized configuration for all subsequent analyses.

\begin{figure}[htbp]
\includegraphics[width=\textwidth]{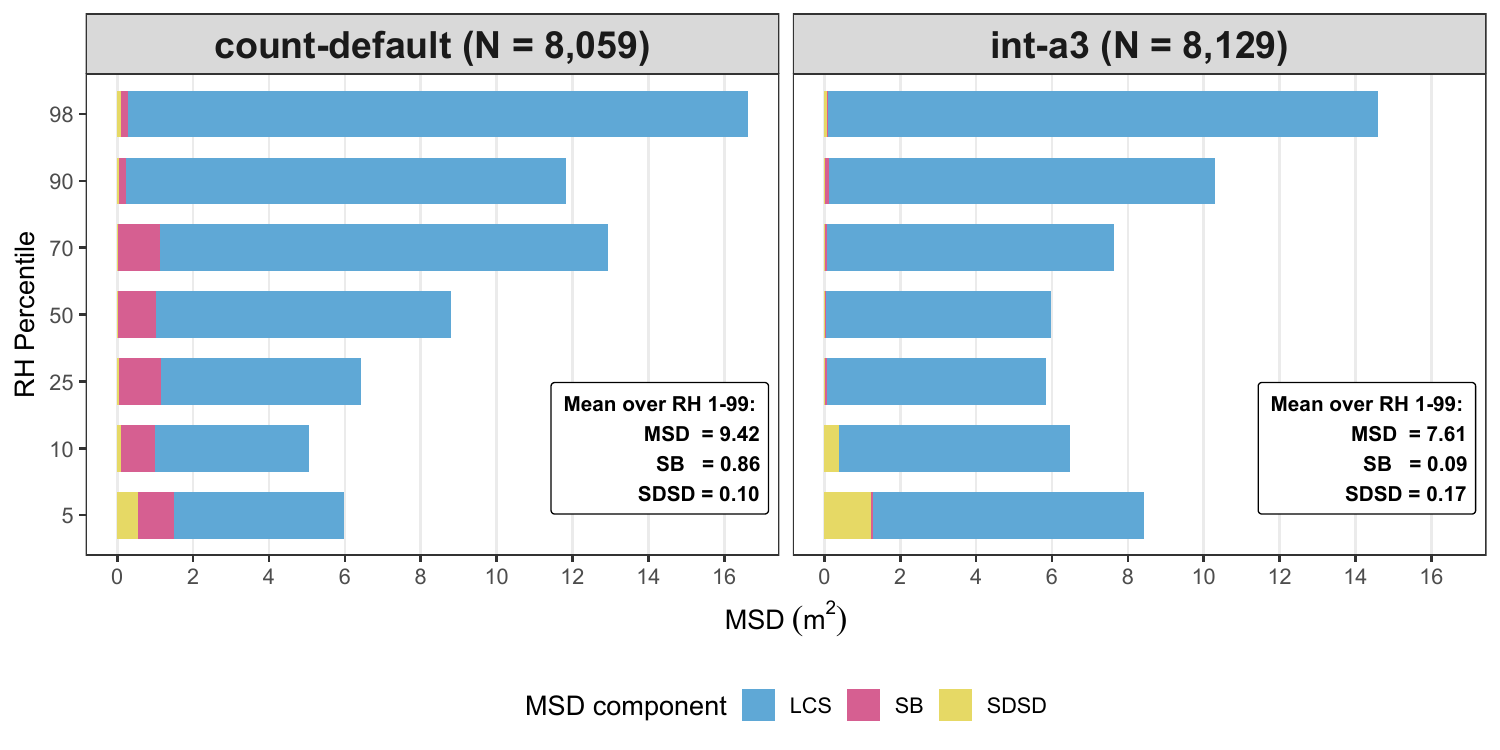}%
\caption{Per-percentile MSD decomposition for selected RH metrics, comparing the \textit{count--default} baseline and the \textit{int--a3} configuration. Bar colours indicate MSD components: LCS (blue), SB (pink), SDSD (yellow). Boxes report mean values aggregated over the full RH1--RH99 profile. Lower values indicate better agreement between simulated and observed GEDI RH profiles. RH metrics are referenced to the ALS-derived ground elevation. $N$ varies mainly across algorithms due to algorithm-specific quality and sensitivity filters,  with only minor variations across weighting schemes.}
\label{fig:msd_stacked}
\end{figure}

To quantify the improvement brought by this optimized setup, we retain the \textit{count--default} configuration as a baseline throughout the remainder of the paper. This baseline mirrors the simulation settings used to calibrate the official GEDI L4A biomass products and is widely adopted in the literature. Comparing it to the \textit{int--a3} configuration allows us to quantify the overall improvement achievable through optimized return weighting and algorithm selection. We illustrate the per-percentile MSD decomposition in Figure~\ref{fig:msd_stacked} for seven RH metrics, contrasting the \textit{count--default} baseline (left panel) with the \textit{int--a3} configuration (right panel). The optimized setup reduces SB at every percentile, with the most pronounced improvements observed in the lower canopy (RH70--RH5), where SB drops to near zero. A residual squared difference between standard deviations (SDSD) contribution persists at the lowest percentiles in both configurations, visible as the yellow bar at RH5, indicating that variability mismatch at the base of the profile is not fully resolved by the change in weighting or algorithm selection alone.

The LCS term remains the dominant component, indicating that residual errors are driven by imperfect co-variation between simulated and observed RH values across footprints. The simulator does not fully capture the spatial variability that drives shot-to-shot differences in real GEDI measurements. Part of this LCS contribution is irreducible by construction, as the stochastic noise present in real GEDI waveforms mechanically degrades the Pearson correlation without substantially affecting the mean or variance of RH distributions, a limitation inherent to any noise-free simulation approach. Bias, therefore, represents the primary actionable metric for evaluating simulator performance.

\subsection{Drivers of Differences Between Simulated and Real GEDI RH}
\label{sec:drivers}

Having identified the optimal simulation configuration (int weighting, a3 algorithm), we now investigate how the residual bias varies with ecological drivers such as forest type and phenological state (leaf-on vs. leaf-off).

\subsubsection{Forest Type}
\label{sec:forest_type}
\begin{figure}[htbp]
\centering
\includegraphics[width=\textwidth]{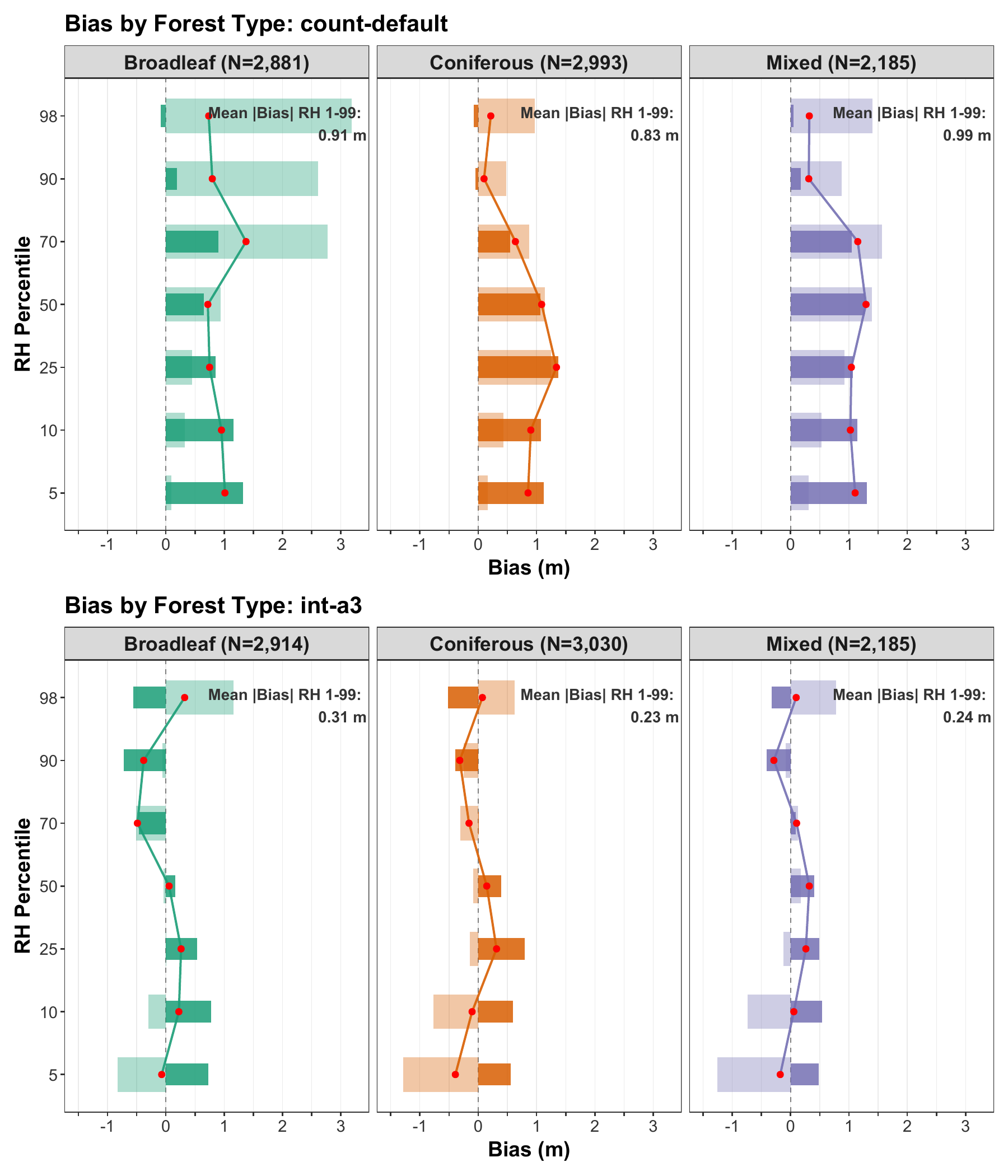}%
\caption{Bias profiles across selected RH percentiles ($\text{bias} = \text{simulated RH} - \text{aligned GEDI RH}$) for broadleaf, coniferous, and mixed forests, comparing the \textit{count--default} baseline (upper row) and the \textit{int--a3} optimized configuration (lower row). For each percentile, wide light bars correspond to low-sensitivity shots (0.90 $ \le \text{sensitivity} \leq$ 0.96) and narrow dark bars to high-sensitivity shots ($\text{sensitivity} >$ 0.96). Red dots indicate the mean bias computed over all shots with $\text{sensitivity} \geq$ 0.90, connected by a line to highlight the overall bias profile. The annotation in each panel reports the mean absolute bias averaged over the full RH1--RH99 range as a global summary measure. All RH metrics are referenced to the ALS-derived ground elevation.}
\label{fig:bias_forest}
\end{figure}

The vertical bias profiles for broadleaf, coniferous, and mixed forests are presented in Figure~\ref{fig:bias_forest} for both the baseline (\textit{count--default}) and optimized (\textit{int--a3}) configurations.

A consistent pattern emerges across all three forest types under the \textit{count--default} baseline: the simulator systematically overestimates RH metrics, with positive biases observed across nearly all percentiles and forest types. The overestimation exceeds 1\,m at specific percentiles that differ by forest type: around RH70 for broadleaf forests, RH50--RH70 for mixed forests, and RH25--RH50 for coniferous forests. This forest-type-specific vertical shift in peak bias may reflect differences in canopy architecture. In coniferous forests, the conical crown shape concentrates foliage in the lower canopy, consistent with the peak bias observed around RH25–RH50, whereas in broadleaf forests the rounded crown places most foliage in the upper canopy, shifting the peak bias toward RH70. In contrast, the upper canopy (RH90--RH98) exhibits consistently low biases across all forest types, with the exception of broadleaf forests where RH90 slightly exceeds this threshold.

A notable divergence is also observed between sensitivity groups. Under the baseline configuration, low-sensitivity shots exhibit substantially larger biases than high-sensitivity shots in the upper canopy (RH70--RH98), with differences exceeding 1.5\,m in some cases. However, this gap narrows progressively toward the lower canopy, and below RH25 the pattern reverses: low-sensitivity shots display equal or lower biases than high-sensitivity shots. This crossing pattern indicates that the baseline configuration does not handle low-sensitivity shots consistently across the profile, motivating the need for an optimized configuration. 

Under the \textit{int--a3} configuration, biases are substantially reduced across all forest types and percentiles, with mean absolute biases of 0.31\,m, 0.23\,m, and 0.24\,m for broadleaf, coniferous, and mixed forests respectively. Comparison with the \textit{count--a3} configuration (see Figure~\ref{fig:bias_count_a3} in \ref{sec:count_frac_a3_appendix}) confirms that most of this improvement stems from intensity weighting rather than algorithm selection alone. Low- and high-sensitivity shots exhibit their smallest bias magnitudes around the mid-canopy (RH50--RH70). Larger biases are found at the profile edges (RH5 and RH98), where the two sensitivity groups display opposite signs. This edge behavior likely reflects differences in signal extent between sensitivity groups: lower-sensitivity shots operate under higher noise conditions, causing the 3$\sigma$ front threshold to define a narrower signal window, which shifts energy distribution toward the profile edges differently than for high-sensitivity shots. These opposing edge biases largely cancel out, keeping the overall mean bias profile close to zero across all forest types. The \textit{int--a3} configuration thus appears robust for simulating GEDI observations across diverse forest types. Although these results reveal consistent improvements, phenology may offer further insight into the mechanisms driving residual biases, as seasonal differences in canopy structure could contribute to the observed patterns under both configurations.

\subsubsection{Phenological Effects}
\label{sec:phenology}

\begin{figure}[htbp]
\centering
\includegraphics[width=\textwidth]{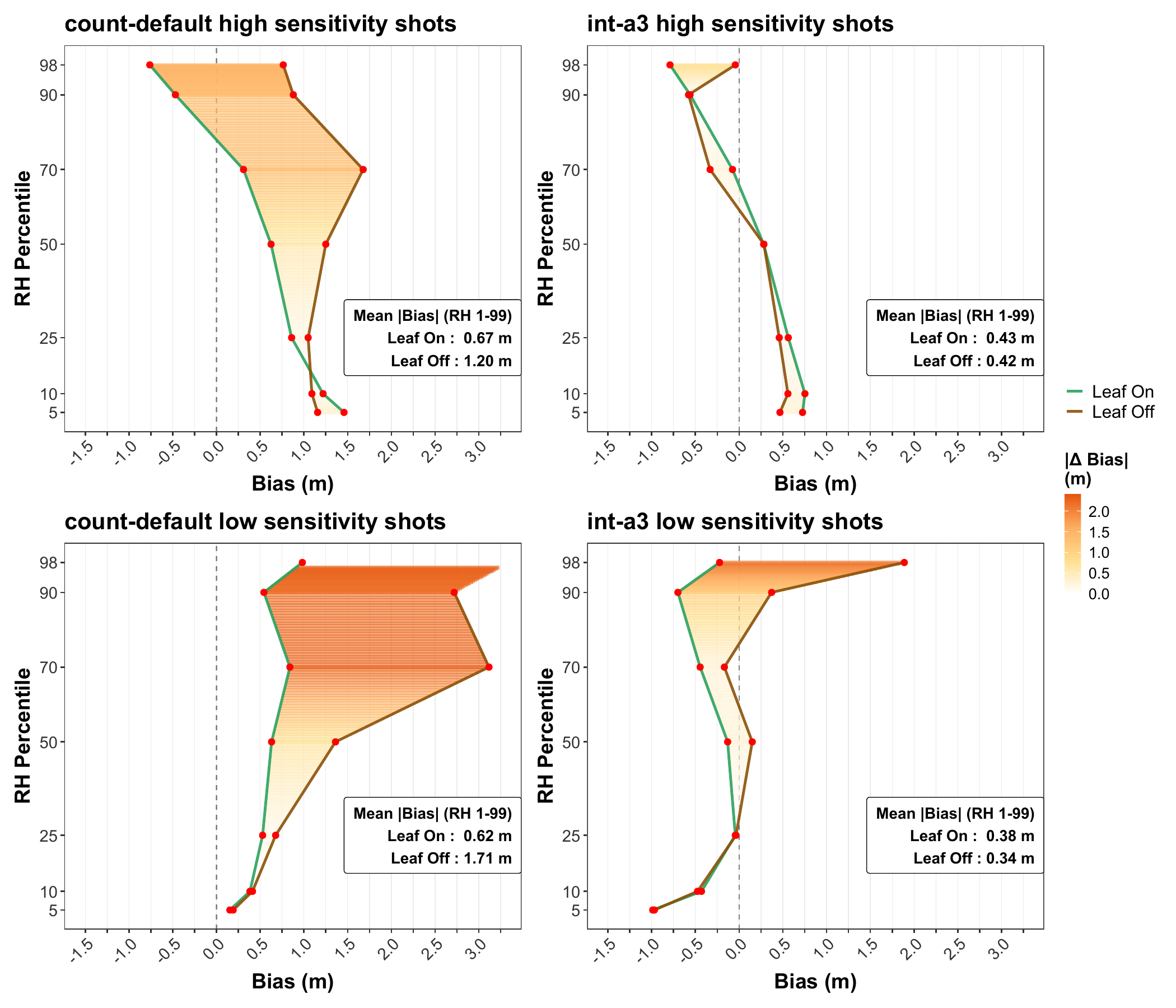}%
\caption{Vertical bias profiles (bias\,=\,mean(simulated RH $-$ aligned GEDI RH)) for leaf-on (green) and leaf-off (brown) conditions under the \textit{count--default} baseline (left) and the \textit{int--a3} optimized configuration (right), restricted to broadleaf and mixed forest types. The shaded area and its colour gradient encode the absolute difference in bias between the two seasons ($|\Delta\text{Bias}|$); darker red indicates larger seasonal divergence. Lines correspond to the bias at the displayed percentiles. Overlaid annotation reports the mean absolute bias averaged over the full RH1--RH99 range, providing an overall summary measure. RH metrics are referenced to the ALS-derived ground elevation.}
\label{fig:bias_phenology}
\end{figure}

To assess phenological effects, we compare leaf-on and leaf-off vertical bias profiles under both configurations (Figure~\ref{fig:bias_phenology}, Table~\ref{tab:phenology_metrics}), focusing on broadleaf and mixed forests. High- and low-sensitivity shots are still separated to provide a complete picture.

In the baseline \textit{count--default} configuration, leaf-off acquisitions are systematically more biased than leaf-on ones. For high-sensitivity shots, the mean absolute bias over the full RH profile is 0.67\,m (leaf-on) and 1.20\,m (leaf-off); for low-sensitivity shots the leaf-off bias reaches 1.71\,m (leaf-on: 0.62\,m). This seasonal imbalance indicates that the overall bias reported in Section \ref{sec:forest_type} partly reflects the baseline's reduced performance under leaf-off conditions.

The \textit{int--a3} configuration substantially reduces the bias under both phenological conditions. For high-sensitivity shots, the mean absolute bias drops to 0.43\,m (leaf-on) and 0.42\,m (leaf-off), a reduction of approximately 65\,\% for leaf-off relative to the baseline. For low-sensitivity shots the improvement is even larger: leaf-on 0.38\,m, leaf-off 0.34\,m (reduction $\sim$80\,\% for leaf-off). The seasonal gap shrinks from 0.53\,m (high-sensitivity) in the baseline to only 0.01\,m with \textit{int--a3}, and a similar collapse is observed for low-sensitivity shots. The only percentile where a strong seasonal divergence persists, even with \textit{int--a3}, is RH98 for low-sensitivity shots: leaf-on bias remains around -0.25\,m whereas leaf-off bias exceeds 1.75\,m. 

To isolate the respective contributions of algorithm selection and return weighting to these seasonal improvements, we also examined the \textit{count--a3} configuration (see Figure~\ref{fig:bias_phenology_count_a3} in \ref{sec:count_frac_a3_appendix}). With \textit{count--a3}, the high-sensitivity leaf-on mean absolute bias is 0.55\,m, already lower than the baseline, but the leaf-off bias remains at 0.91\,m, still twice as large as the 0.42\,m obtained with \textit{int--a3}. Because both configurations are referenced to the same ALS-derived ground elevation and use the same L2A processing algorithm (a3) for RH computation, this comparison cleanly isolates the effect of return weighting: the \texttt{int} method is responsible for the largest part of the leaf-off improvement, whereas algorithm selection alone cannot resolve the seasonal bias.

\begin{table}[htbp]
\caption{Per-Percentile Bias, RMSE, and R$^2$ Between Simulated and Aligned GEDI RH Metrics for Leaf-On and Leaf-Off Conditions, Stratified by Configuration and Sensitivity Group}
\centering
\small
\begin{tabular}{l l rrr rrr}
\toprule
& & \multicolumn{3}{c}{\textbf{Leaf-on}} & \multicolumn{3}{c}{\textbf{Leaf-off}} \\
\cmidrule(lr){3-5} \cmidrule(lr){6-8}
\textbf{Configuration} & \textbf{RH} & \textbf{Bias (m)} & \textbf{RMSE (m)} & \textbf{R$^2$} & \textbf{Bias (m)} & \textbf{RMSE (m)} & \textbf{R$^2$} \\
\midrule
\multicolumn{8}{l}{\shortstack[l]{\textit{count--default} \\ (high sensitivity)}} \\
& RH25 &  0.86 & 2.04 & 0.82 & 1.05 & 2.33 & 0.82 \\
& RH50 &  0.62 & 2.45 & 0.81 & 1.25 & 3.06 & 0.80 \\
& RH70 &  0.31 & 2.77 & 0.80 & 1.68 & 4.19 & 0.69 \\
& RH98 & -0.76 & 2.97 & 0.83 & 0.76 & 4.35 & 0.71 \\
\midrule
\multicolumn{8}{l}{\shortstack[l]{\textit{count--default} \\ (low sensitivity)}} \\
& RH25 &  0.53 & 1.64 & 0.70 & 0.68 & 2.00 & 0.71 \\
& RH50 &  0.63 & 2.37 & 0.70 & 1.36 & 3.33 & 0.61 \\
& RH70 &  0.84 & 3.27 & 0.65 & 3.12 & 5.89 & 0.20 \\
& RH98 &  0.98 & 4.19 & 0.65 & 3.35 & 6.70 & 0.36 \\
\midrule
\multicolumn{8}{l}{\shortstack[l]{\textit{int--a3} \\ (high sensitivity)}} \\
& RH25 &  0.56 & 1.97 & 0.83 & 0.46 & 1.98 & 0.87 \\
& RH50 &  0.28 & 2.15 & 0.85 & 0.28 & 2.03 & 0.91 \\
& RH70 & -0.08 & 2.46 & 0.84 & -0.33 & 2.76 & 0.85 \\
& RH98 & -0.79 & 2.98 & 0.83 & -0.04 & 3.72 & 0.77 \\
\midrule
\multicolumn{8}{l}{\shortstack[l]{\textit{int--a3} \\ (low sensitivity)}} \\
& RH25 & -0.04 & 1.83 & 0.87 & -0.04 & 2.27 & 0.84 \\
& RH50 & -0.13 & 1.81 & 0.90 & 0.15 & 2.58 & 0.87 \\
& RH70 & -0.45 & 2.28 & 0.87 & -0.17 & 3.53 & 0.80 \\
& RH98 & -0.22 & 3.35 & 0.78 & 1.89 & 5.40 & 0.63 \\
\bottomrule
\end{tabular}
\par\smallskip
\begin{minipage}{\linewidth}
\footnotesize \textbf{Note.} Broadleaf and mixed forests only. Sample sizes: \textit{count--default} high sensitivity: $n_{\text{on}}=2{,}024$, $n_{\text{off}}=1{,}882$; low sensitivity: $n_{\text{on}}=405$, $n_{\text{off}}=755$. \textit{int--a3} high sensitivity: $n_{\text{on}}=1{,}490$, $n_{\text{off}}=1{,}281$; low sensitivity: $n_{\text{on}}=954$, $n_{\text{off}}=1{,}374$.
\end{minipage}
\label{tab:phenology_metrics}
\end{table}

Overall, the \textit{int--a3} configuration makes leaf-off acquisitions as reliable as leaf-on ones across most of the vertical profile, except for RH98 in low-sensitivity shots (Table~\ref{tab:phenology_metrics}), which should be treated with caution.

\clearpage
\section{Discussion and Implications for Users}

\subsection{Primary Driver of Bias}

\subsubsection{Ground Detection: An Overlooked Source of Bias}

Our evaluation of GEDI ground elevation highlights a key but undocumented source of uncertainty when calibrating models intended for use with GEDI products: the choice of L2A processing algorithm. Among the six algorithms, a3 shows the smallest overall residual relative to ALS ground (bias +0.04\,m), outperforming both the default configuration (bias -0.82\,m) and the commonly used a1 and a2 (biases +1.64\,m and -0.79\,m respectively). This near-zero bias comes at the cost of a slightly higher RMSE (4.54\,m vs. 4.07\,m for the default), a trade-off that is acceptable here because a systematic ground-detection offset propagates directly into all RH metrics, whereas the additional dispersion can be partly averaged out in statistical mapping frameworks. This improvement arises from the specific combination of a wide smoothing window for noise identification with a narrow denoising window for zero-crossing, which preserves finer features of the ground return and avoids the over-flattening introduced by more aggressive filters.

The advantage of a3 is, however, mainly driven by high-sensitivity shots. In the low-sensitivity class ($0.90 \le \text{sens} \le 0.96$), the default configuration achieves a smaller bias (-0.14\,m) and lower RMSE (2.70\,m) than a3 (+0.70\,m, 4.70\,m). This reflects a difference in how each algorithm estimates sensitivity. The default configuration assigns far fewer shots to the low-sensitivity category (1,988 vs. 3,893 for a3), meaning its low-sensitivity subset is more restrictive and contains on average fewer challenging waveforms. On the other hand, a3 classifies a larger share of footprints as low-sensitivity, including many borderline cases that the default algorithm pushes into the high-sensitivity class. As a result, the low-sensitivity bias of the default appears smaller, but the high-sensitivity bias of the default (-1.04\,m) is much worse than that of a3 (-0.57\,m). The near-zero overall bias of a3 thus reflects a more balanced classification and a conservative back-threshold that avoid the large systematic offset seen in the default's high-sensitivity group.

Because RH metrics are expressed relative to the detected ground, any algorithm-dependent bias propagates directly into RH errors. The widespread use of the default GEDI configuration, which in our dataset relies exclusively on a1 or a2, introduces a systematic and unreported shift in RH metrics. This bias is overlooked when users rely solely on the default product, since alternative algorithms are rarely evaluated. Our results demonstrate that a spaceborne LiDAR sensor can achieve near-unbiased ground detection relative to ALS when an appropriate algorithm is selected, and suggest that future GEDI reprocessing efforts could reduce this uncertainty simply by revisiting algorithm selection criteria. This could directly enhance biomass prediction accuracy, as models calibrated on simulated RH profiles referenced to the ALS ground, would be applied to real GEDI RH metrics referenced to a more accurate detected ground, reducing the systematic offset between calibration and inference.

\subsubsection{Return-Weighting Strategies}

Beyond ground detection, a3 also yields low squared bias on the RH metrics themselves (Table~\ref{tab:msd_decomposition_full}), suggesting that its denoising parameters preserve the vertical energy distribution more faithfully than the default algorithms. However, the most substantial reduction in RH bias arises not from algorithm choice alone but from switching the return-weighting strategy from \texttt{count} to \texttt{int}. We therefore examine the mechanisms by which intensity-based weighting improves waveform simulation, and show that this improvement is robust across sensitivity classes, forest types, and phenological conditions.

The \texttt{count}, \texttt{frac}, and \texttt{int} weighting options differ in how they translate discrete ALS returns into a continuous waveform. Under \texttt{count}, every return contributes equally to the vertical profile, thereby implicitly assuming that all intercepted targets share identical characteristics. In reality, the energy scattered back toward the sensor depends on the reflectance, orientation, and fractional coverage of the intercepting surface. The \texttt{int} method approximates this behaviour by weighting each return by its recorded ALS intensity, while \texttt{frac} distributes the outgoing energy equally among all returns derived from the same pulse.

Our results show that \texttt{int} consistently yields the smallest bias and MSD across all L2A algorithms and sensitivity classes (Tables~\ref{tab:msd_decomposition_low_sensitivity_N_inline}--\ref{tab:msd_decomposition_high_sensitivity_N_inline} in \ref{sec:low_high_appendix}). This indicates that a high-density ALS point cloud acquired at 1064\,nm captures physically meaningful variations in target reflectance, and that \texttt{int} weighting translates these variations into a more faithful representation of the scattering surfaces that GEDI actually detects. Bare soil produces a concentrated, high-amplitude ground return due to the absence of volume scattering, whereas vegetation returns are spread vertically across multiple canopy layers. By giving proportionally more weight to these distinctive ground returns, \texttt{int} reconstructs a ground peak whose amplitude and width more closely match the real GEDI ground return. Within the canopy, intensity also encodes the effective intercepted area of each scattering element: dense foliage clusters and large branches return higher intensities than isolated fine branches or sparse leaf edges, so \texttt{int} naturally concentrates energy in the dominant canopy layers rather than distributing it uniformly across all returns. This produces a vertical energy distribution that reflects the actual backscatter profile recorded by GEDI.

Under \texttt{count}, the ground peak is underweighted relative to the canopy signal. The simulated cumulative energy therefore leaves the ground zone with a small share of the total energy, and each percentile is reached higher in the profile. In the real GEDI waveform, the prominent ground return holds a large share of the energy near the ground, so the same percentiles are reached lower, which explains the systematic positive bias of the simulated RH (\ref{sec:waveformsplots}) under both seasons (see Figures~\ref{fig:bias_forest}--\ref{fig:bias_phenology}). Regarding the canopy signal, \texttt{frac}, by distributing pulse energy among multiple returns from the same shot, partially reduces the overestimation of upper percentiles compared to \texttt{count}, which assigns full weight to every return regardless of pulse origin and therefore strongly inflates the canopy contribution. The \texttt{int} method outperforms both by reflecting the true radiometric contrast between surfaces. Visual inspection of simulated waveforms (\ref{sec:waveformsplots}) confirms that \texttt{int} produces both a ground return and a canopy signal that align more closely with the observed GEDI waveform, directly reducing these residual biases.

The performance of \texttt{int} is further supported by its pairing with the a3 algorithm, whose properties complement the weighting strategy. Unlike the default algorithms, a3 applies a less aggressive denoising window, which better preserves the fine structure of the energy distribution. This closer correspondence between the a3 cumulative energy profile and the noise-free simulated waveform explains why the \textit{int--a3} combination achieves both the lowest ground detection bias and the lowest RH bias simultaneously. A key consequence of this improvement is the near-disappearance of the seasonal bias gap. With \textit{int--a3}, high-sensitivity leaf-off acquisitions achieve a mean absolute bias of 0.42\,m, almost identical to the 0.43\,m obtained for leaf-on conditions, whereas the baseline \textit{count--default} configuration produced a bias of 1.20\,m for leaf-off and 0.67\,m for leaf-on. The \textit{int--a3} configuration thus removes most of the seasonal asymmetry that affected the baseline, demonstrating that the simulator can emulate leaf-off GEDI waveforms as faithfully as leaf-on ones when return weights reflect the radiometric properties of the intercepted surfaces. 

The only notable exception is the upper percentiles (RH98) in leaf-off, low-sensitivity conditions, where all weighting methods produce a positive bias of approximately 2\,m. In these conditions, the weak returns from bare upper branches fall below the $3\sigma$ front threshold used by GEDI to define the start of the usable signal, causing the detected canopy top to be placed lower than the true top. The noise-free simulated waveform captures all ALS reflections from these sparse bare branches regardless of their energy, and therefore overestimates RH98 relative to the observed metric. This is not an inherent limitation of the simulator but rather a consequence of GEDI's noise thresholding: high-sensitivity shots (for any phenological state) and low-sensitivity leaf-on shots perform well, confirming that the issue arises specifically from the combination of low signal-to-noise ratio and sparse leaf-off canopy. RH98 in leaf-off, low-sensitivity shots should therefore be treated with caution or excluded from downstream analyses.

Our results support the use of \texttt{int} weighting for large-scale GEDI simulation studies, particularly when leaf-off ALS data must be incorporated to maximize the calibration sample size. This implies that biomass models could be calibrated on simulated metrics derived from leaf-on or leaf-off ALS acquisitions, substantially increasing the pool of ALS data available for calibration. In addition, it means that the inference dataset available for large-area mapping could be extended to real leaf-off GEDI observations. Future developments of the simulator could explore hybrid strategies combining \texttt{int} and \texttt{frac} to better handle multiple returns from a single pulse, or incorporate scan-angle-dependent intensity corrections to further reduce the small residual biases observed.

\subsubsection{Low and High Sensitivity Shots}

At the level of the full RH profile (RH1--RH99), the two sensitivity groups perform comparably under the \textit{int--a3} configuration: low-sensitivity shots yield a mean SB of 0.27 (Table~\ref{tab:msd_decomposition_low_sensitivity_N_inline} in \ref{sec:low_high_appendix}) compared to 0.21 for high-sensitivity shots (Table~\ref{tab:msd_decomposition_high_sensitivity_N_inline} in \ref{sec:low_high_appendix}), indicating that low-sensitivity shots are not systematically more biased than their higher-energy counterparts. The lower SNR of coverage beams mainly degrades precision rather than accuracy: under \textit{int--a3} (Table~\ref{tab:msd_decomposition_low_sensitivity_N_inline}-\ref{tab:msd_decomposition_high_sensitivity_N_inline} in  \ref{sec:low_high_appendix}), the higher MSD of low-sensitivity shots (8.78 vs 6.53\,m$^2$) is driven almost entirely by the larger LCS term (8.13 vs 6.27), reflecting increased shot-to-shot dispersion rather than a systematic offset. This distinction matters for large-scale biomass mapping: including low-sensitivity shots, after appropriate averaging or within statistical frameworks, could substantially increase effective sampling density without introducing additional systematic error.

However, the reliability of low-sensitivity RH metrics depends on which percentiles are used. The most robust subset spans RH25 to RH90 (see Figure \ref{fig:bias_phenology}), as these values are sufficiently far from the signal boundaries where threshold instability is highest. The marginal percentiles RH5 and RH98 remain problematic, particularly under leaf-off conditions, as shown in Figure~\ref{fig:bias_phenology}. Making these percentiles reliable for low-sensitivity shots would require further work explicitly accounting for noise in the simulations, for instance by injecting synthetic noise into simulated waveforms to better reproduce the thresholding behaviour of real GEDI observations, or by developing a dedicated L2A processing algorithm that applies more conservative signal extent thresholds for low-sensitivity waveforms.

Regarding spatial representativeness, our study did not explicitly investigate the performance of coverage beams across topographic gradients. Because low-sensitivity shots are more susceptible to degraded ground detection in complex terrain, they may be underrepresented in the most mountainous areas, where challenging waveform conditions cause more shots to fall below quality thresholds. Users should therefore be aware that, while coverage beams perform comparably to full-power beams across the range of conditions sampled here, their spatial distribution may not be uniform across all terrain types. We nonetheless encourage their inclusion in large-scale studies, while recommending that potential spatial sampling imbalances across topographic gradients be monitored when interpreting results.

\subsubsection{Generalisability and Recommendations for Simulator Users}

Our conclusions should be interpreted in the context of the specific conditions under which this validation was conducted, but several characteristics of our dataset support a broader applicability of our findings to other European temperate forests. 

The original validation of the simulator \cite{Hancock} focused on tropical forests and North American temperate forests acquired with lower-density ALS point clouds, and found \texttt{count} to be the most accurate weighting strategy. Our results differ, with \texttt{int} outperforming \texttt{count} across all configurations. This difference likely reflects not only the higher point density of our ALS data, but also key sensor characteristics. The North American sites in the original validation were acquired with older discrete-return sensors operating at lower point densities, whereas both the RIEGL VQ-480 and the Galaxy T2000 used here are full-waveform sensors capable of recording a larger number of discrete returns, providing richer radiometric sampling within each pulse. Furthermore, the original Gabon validation relied on a sensor operating at 1550 nm, a wavelength at which the reflectance contrast between bare soil and vegetation differs substantially from that at 1064 nm, potentially affecting the relative performance of intensity-based weighting. 

These sensor-level differences may provide a mechanistic explanation for why \texttt{int} weighting outperforms \texttt{count} in our dataset. Modern full-waveform sensors likely capture physically meaningful radiometric variations across canopy layers and between vegetation and bare soil. Intensity-based weighting may translate these variations into a more faithful representation of the backscatter profile that GEDI actually detects. The RIEGL VQ-480 and Galaxy T2000 are not specific to the French program: they represent the current generation of airborne LiDAR systems increasingly adopted for national-scale ALS acquisitions across Europe. Given the widespread adoption of these modern sensors, our recommendation to use \texttt{int} weighting may be relevant to a large and growing fraction of the European LiDAR community.

Beyond sensor characteristics, the use of the Corine Land Cover classification as our forest stratification framework further supports the transferability of our results. CLC applies consistent definitions of broadleaf, coniferous, and mixed forest classes across the entire European continent, meaning that the forest types evaluated here are directly comparable to those found in other European countries. Metropolitan France encompasses a wide range of temperate forest conditions, from Atlantic oak forests in the west to montane conifer stands in the Alps and Pyrenees, that represent a substantial portion of the structural variability found in European temperate forests. We therefore suggest that our results constitute a relevant baseline for other European contexts where similar CLC forest classes and comparable ALS sensor configurations are used, while acknowledging that ecosystems at the margins of this range, such as boreal forests in Scandinavia or dense Mediterranean woodlands, may require independent validation.

These two arguments, comparable forest types and converging sensor standards, suggest that the \textit{int--a3} configuration may constitute a relevant starting point for GEDI simulation studies in European temperate forests using modern high-density ALS data. We still encourage users working in other biomes or with older sensor configurations to conduct their own validation against real GEDI observations before treating simulated metrics as an unbiased reference, as the optimal weighting strategy may differ depending on point density, wavelength, and forest structural characteristics.

\section{Conclusion}

Our study confirms the value of the simulator for understanding GEDI's sensitivity to forest structure and for supporting the calibration of biomass models. Using a dataset of approximately 8,100 collocated simulated and real GEDI waveform pairs across French forests, we show that, for high-density point clouds ($\geq 10$\,points\,m$^{2}$) acquired with up-to-date full-waveform ALS systems, the baseline \texttt{count} weighting, combined with the default L2A algorithm, overestimates most of the RH metrics as it treats each ALS return as an equal contributor regardless of its radiometric properties. This contrasts with the initial validation of the simulator \cite{Hancock}, where \texttt{count} was the most accurate option. The difference likely reflects ALS sensor characteristics, point density and wavelength, although these factors could not be tested independently here. The \texttt{int} option, which weights each return by its recorded intensity, provides a physically more realistic representation of the scattering surfaces that GEDI actually detects. Combined with the a3 L2A algorithm, which achieves near-unbiased ground detection (bias +0.04\,m versus  -0.82\,m for the default), this configuration reduces the mean absolute bias across the full RH profile from 0.67\,m to  0.43\,m in leaf-on conditions and from 1.20\,m to 0.42\,m in leaf-off conditions for high-sensitivity shots.

With \textit{int--a3}, the seasonal bias gap nearly disappears, making leaf-off ALS acquisitions and leaf-off GEDI shots as reliable as leaf-on ones in terms of mean bias for simulating RH metrics, except for RH98 in low-sensitivity shots. This opens the way to using leaf-off data, currently excluded from standard GEDI biomass products, for biomass calibration and large-scale inference, pending validation on AGBD models. Low-sensitivity shots (down to a sensitivity of 0.9) do not show a larger systematic bias than high-sensitivity ones: their lower signal-to-noise ratio mainly increases shot-to-shot dispersion, and they can be included in large-scale studies, excluding RH98 under leaf-off conditions and the lowest percentiles (RH5 and below). Potential spatial sampling imbalances across topographic gradients should nonetheless be monitored when coverage beams are included in statistical mapping frameworks. By controlling these biases through appropriate weighting and algorithm selection, the usable GEDI dataset for large-scale forest monitoring and biomass estimation is considerably expanded.

\subsection{Perspectives}

Building on the validated \textit{int--a3} configuration, the next step will be to simulate unbiased GEDI-like metrics at National Forest Inventory plot locations across France to calibrate local aboveground biomass density models. The findings of this study open the way for testing whether incorporating leaf-off ALS acquisitions for calibration, and leaf-off and low-sensitivity GEDI shots for inference, can improve biomass estimates and expand spatial coverage relative to standard approaches. The paired dataset of real and simulated GEDI waveforms produced here is openly available \cite{behloul_2026_dataset} and may serve as a benchmark for transfer learning and domain adaptation studies targeting other European temperate forest regions.


\section*{Open Research Section}
GEDI L1B and L2A products (V2) were downloaded from NASA Earthdata Search (\url{https://search.earthdata.nasa.gov}). The French national LiDAR HD point clouds are distributed by the Institut national de l'information géographique et forestière (IGN) via the Geoplateforme (\url{https://geoservices.ign.fr/lidarhd}). The Corine Land Cover 2018 raster dataset \cite{clc2018_raster} is available from the Copernicus Land Monitoring Service. The GEDI simulator code is available at \url{https://bitbucket.org/StevenHancock/gedisimulator}. The paired real and simulated GEDI dataset produced in this study is available at Zenodo (\url{https://doi.org/10.5281/zenodo.23402593}) \cite{behloul_2026_dataset}. Analysis scripts are available on GitHub at \url{https://github.com/Selimgit/GEDI-LHD_simulations} and are archived on Zenodo at \url{https://doi.org/10.5281/zenodo.23601741} \cite{behloul_2026_code}.


\section*{Conflict of Interest Disclosure}
The authors declare there are no conflicts of interest for this manuscript.


\acknowledgments
This work was supported by project AI4FI as part of the French France 2030 program ``Initiative d'Excellence Lorraine (LUE)'' (ANR-15-IDEX-04-LUE). This work was also supported by the ANR funded research project AI4Forest (ANR-22-FAI1-0002); the project ALAMOD of the exploratory research program FairCarboN (ANR-22-PEXF-0002); the project Monitor of the research program PEPR FORESTT (ANR-24-PEFO-0003); and by the CLAND Convergence Institute funded by the French National Research Agency (ANR) under grant ANR-16-CONV-0003. High Performance Computing resources were partially provided by the EXPLOR centre hosted by the University of Lorraine (Project LIFEXPLOR 2025EXTXX3744).

\bibliography{ref}

\clearpage
\appendix

\section{Real and Simulated Waveforms}
\label{sec:waveformsplots}

\begin{figure}[htbp]
\noindent\includegraphics[width=0.92\textwidth]{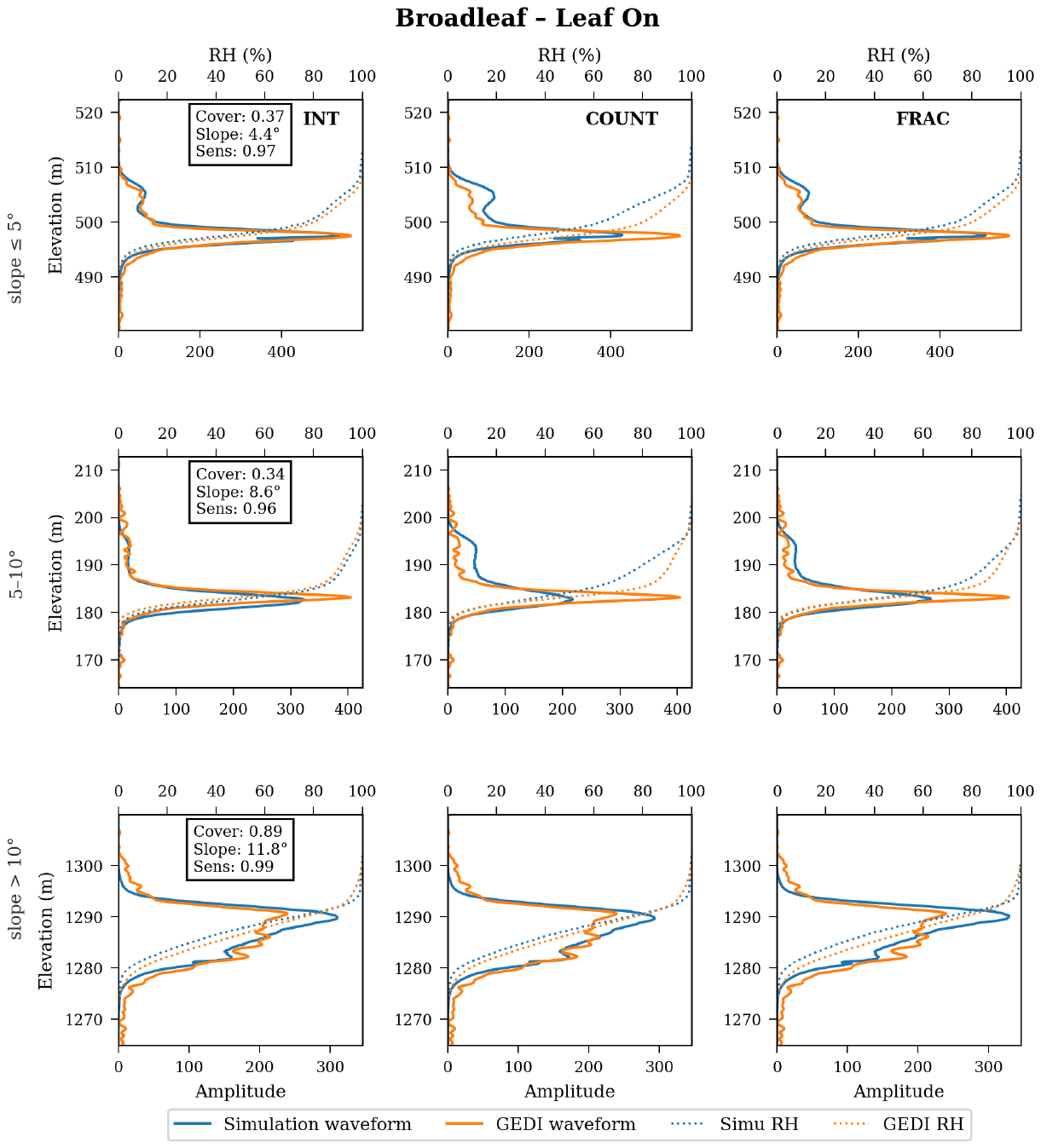}
\caption{Simulated and real GEDI waveforms for a broadleaf canopy in leaf-on conditions. Each column corresponds to a different weighting method. The overlaid RH profiles are: the GEDI default configuration profile and the simulated profile, both derived from ALS ground.}
\label{fig:leafon}
\end{figure}
\clearpage
\begin{figure}[htbp]
\noindent\includegraphics[width=0.92\textwidth]{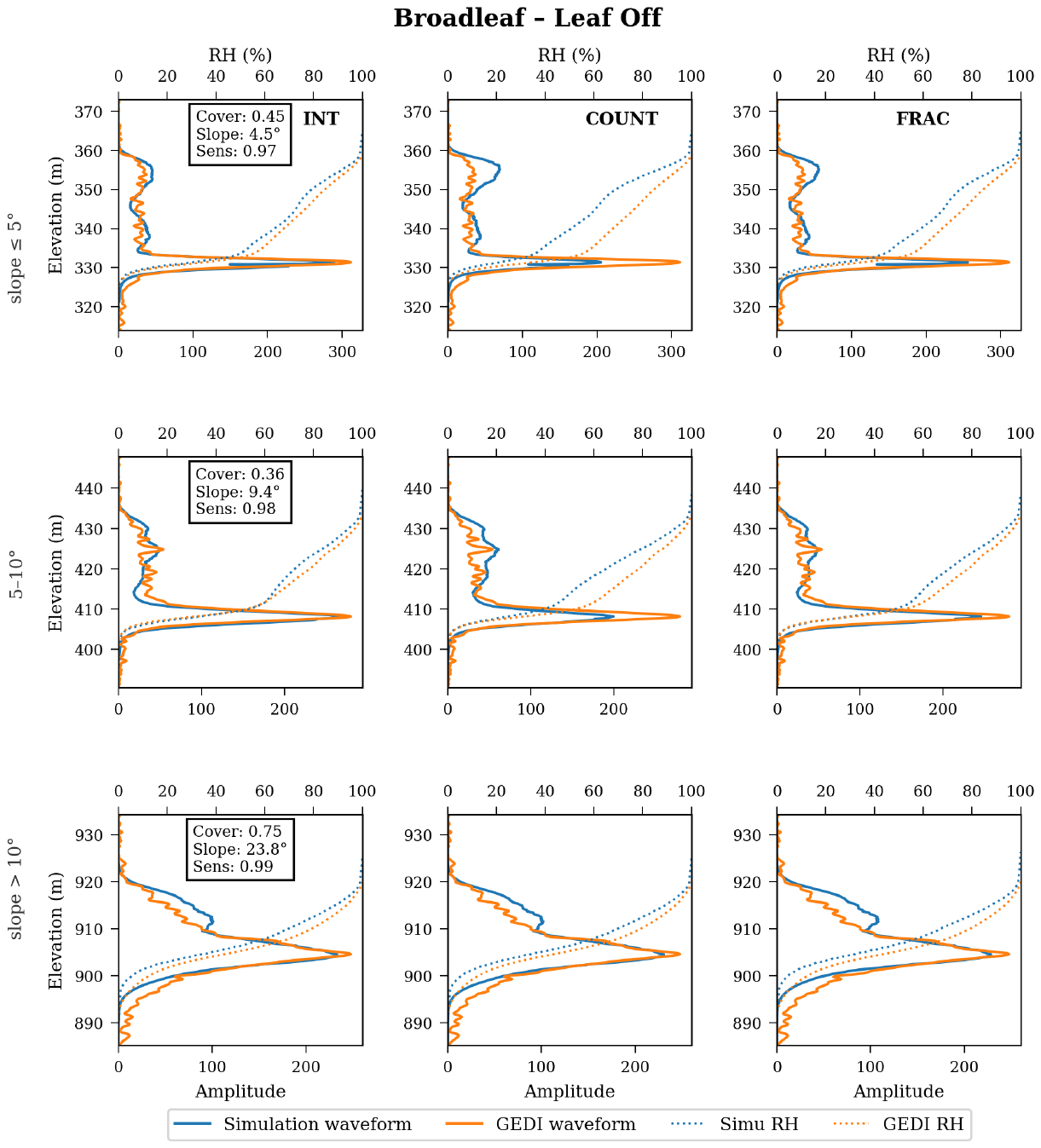}
\caption{Simulated and real GEDI waveforms for a broadleaf canopy in leaf-off conditions. Each column represents a distinct weighting method. The overlaid RH profiles are: the GEDI default configuration profile and the simulated profile, both derived from ALS ground.}
\label{fig:leafoff}
\end{figure}
\clearpage
\begin{figure}[htbp]
\noindent\includegraphics[width=0.92\textwidth]{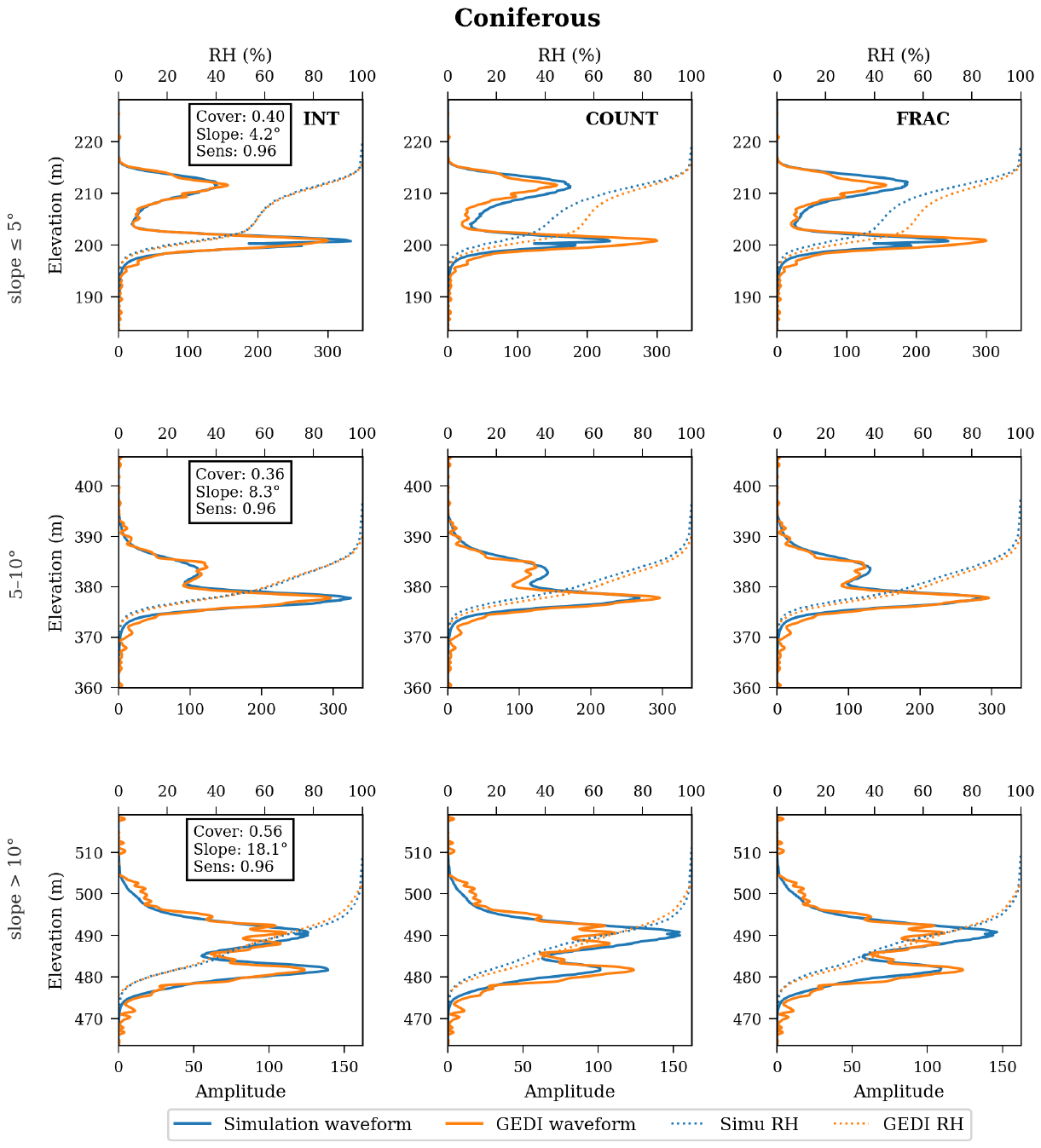}
\caption{Simulated and real GEDI waveforms for a coniferous canopy. Each column illustrates a different weighting method. The overlaid RH profiles are: the GEDI default configuration profile and the simulated profile, both derived from ALS ground.}
\label{fig:coniferous}
\end{figure}

\clearpage
\section{Low and High Sensitivity Shots Analysis}

\begin{table}[htbp]
\caption{Mean MSD Decomposition (SB, SDSD, LCS, MSD) for Low Sensitivity Shots (Sensitivity Between 0.9 and 0.96)}
\centering
\fontsize{9}{11}\selectfont
\begin{tabular}{l c c c c c c c}
\toprule
\textbf{Metric} & \textbf{default} & \textbf{a1} & \textbf{a2} & \textbf{a3} & \textbf{a4} & \textbf{a5} & \textbf{a6} \\
\midrule
SB count   & \textbf{1.52} & 0.53 & 1.83 & 0.64 & 3.48 & 2.81 & 1.30 \\
SB frac    & 0.53 & 0.23 & 0.97 & 0.31 & 2.59 & 1.88 & 0.62 \\
SB int     & 0.17 & 0.19 & 0.40 & \textbf{0.27} & 1.67 & 1.13 & 0.20 \\
\midrule
SDSD count & \textbf{0.16} & 0.48 & 0.24 & 0.61 & 0.52 & 0.25 & 0.20 \\
SDSD frac  & 0.22 & 0.29 & 0.37 & 0.38 & 0.31 & 0.35 & 0.21 \\
SDSD int   & 0.18 & 0.29 & 0.33 & \textbf{0.38} & 0.29 & 0.30 & 0.20 \\
\midrule
LCS count  & \textbf{10.39} & 9.64 & 10.08 & 10.28 & 15.32 & 11.55 & 9.84 \\
LCS frac   & 8.57 & 8.06 & 8.51 & 8.58 & 12.87 & 10.45 & 8.20 \\
LCS int    & 8.25 & 7.70 & 7.94 & \textbf{8.13} & 11.66 & 10.01 & 7.63 \\
\midrule
MSD count  & \textbf{12.07} & 10.65 & 12.16 & 11.54 & 19.31 & 14.61 & 11.34 \\
MSD frac   & 9.33 & 8.59 & 9.85 & 9.28 & 15.77 & 12.68 & 9.03 \\
MSD int    & 8.60 & 8.18 & 8.67 & \textbf{8.78} & 13.63 & 11.44 & 8.04 \\
\bottomrule
\end{tabular}
\par\smallskip
\begin{minipage}{\linewidth}
\footnotesize \textbf{Note.} All values in m$^2$. RH metrics are referenced to the ALS-derived ground elevation. Results are for $0.90 \le \text{sensitivity} \le 0.96$. Sample sizes are reported for each algorithm using the \texttt{count} weighting: default (1,988), a1 (4,239), a2 (800), a3 (3,893), a4 (4,239), a5 (82), a6 (2,358). $N$ varies mainly across algorithms due to algorithm-specific quality and sensitivity filters,  with only minor variations across weighting schemes.
\end{minipage}
\label{tab:msd_decomposition_low_sensitivity_N_inline}
\end{table}

\clearpage
\label{sec:low_high_appendix}
\begin{table}[htbp]
\caption{Mean MSD Decomposition (SB, SDSD, LCS, MSD) for High Sensitivity Shots (Sensitivity Above 0.96)}
\centering
\fontsize{9}{11}\selectfont
\begin{tabular}{l c c c c c c c}
\toprule
\textbf{Metric} & \textbf{default} & \textbf{a1} & \textbf{a2} & \textbf{a3} & \textbf{a4} & \textbf{a5} & \textbf{a6} \\
\midrule
SB count   & \textbf{0.82} & 0.53 & 0.99 & 0.53 & 0.78 & 1.50 & 0.68 \\
SB frac    & 0.74 & 0.46 & 0.73 & 0.40 & 0.64 & 1.17 & 0.51 \\
SB int     & 0.44 & 0.28 & 0.40 & \textbf{0.21} & 0.28 & 0.73 & 0.29 \\
\midrule
SDSD count & \textbf{0.10} & 0.12 & 0.10 & 0.12 & 0.13 & 0.13 & 0.13 \\
SDSD frac  & 0.09 & 0.07 & 0.11 & 0.07 & 0.07 & 0.17 & 0.09 \\
SDSD int   & 0.08 & 0.06 & 0.09 & \textbf{0.06} & 0.06 & 0.15 & 0.07 \\
\midrule
LCS count  & \textbf{7.64} & 7.36 & 8.42 & 7.52 & 8.01 & 8.74 & 8.00 \\
LCS frac   & 6.68 & 6.44 & 7.15 & 6.54 & 6.84 & 7.44 & 6.85 \\
LCS int    & 6.30 & 6.10 & 6.75 & \textbf{6.27} & 6.33 & 7.00 & 6.53 \\
\midrule
MSD count  & \textbf{8.56} & 8.02 & 9.51 & 8.17 & 8.92 & 10.37 & 8.80 \\
MSD frac   & 7.50 & 6.97 & 7.99 & 7.01 & 7.56 & 8.78 & 7.45 \\
MSD int    & 6.82 & 6.43 & 7.24 & \textbf{6.53} & 6.67 & 7.88 & 6.89 \\
\bottomrule
\end{tabular}
\par\smallskip
\begin{minipage}{\linewidth}
\footnotesize \textbf{Note.} All values in m$^2$. RH metrics are referenced to the ALS-derived ground elevation. Results are for $0.96 < \text{sens}$. Sample sizes are reported for each algorithm using the \texttt{count} weighting: default (6,071), a1 (3,508), a2 (7,435), a3 (4,246), a4 (3,508), a5 (8,141), a6 (5,873). $N$ varies mainly across algorithms due to algorithm-specific quality and sensitivity filters,  with only minor variations across weighting schemes.
\end{minipage}
\label{tab:msd_decomposition_high_sensitivity_N_inline}
\end{table}

\clearpage
\section{\texttt{count} and \texttt{frac} results with a3 L2A algorithm}
\label{sec:count_frac_a3_appendix}
\begin{figure}[htbp]
\centering
\includegraphics[width=\textwidth]{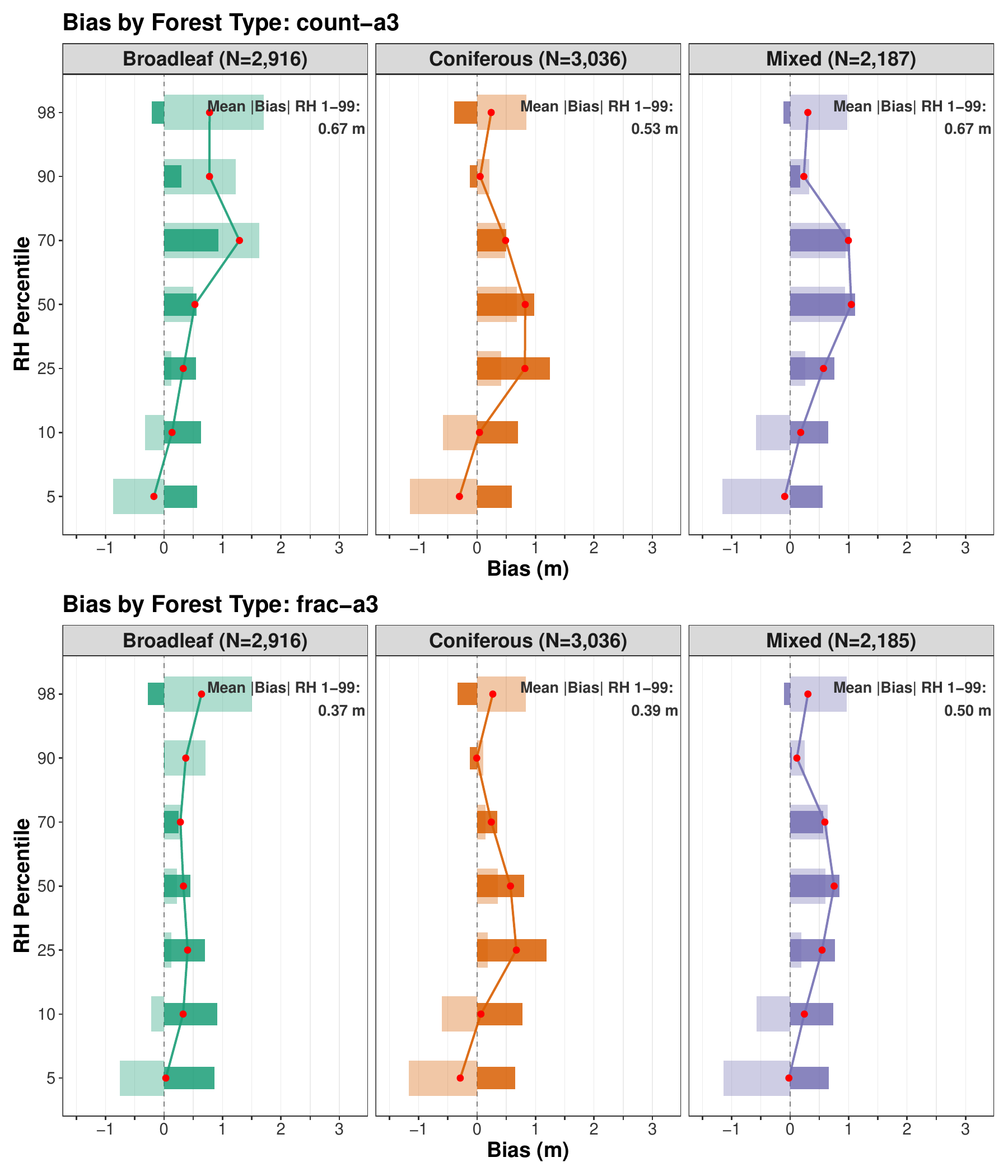}%
\caption{Bias profiles across selected RH percentiles ($\text{bias} = \text{simulated RH} - \text{aligned GEDI RH}$) for broadleaf, coniferous, and mixed forests, comparing the \textit{count--a3} baseline (upper row) and the \textit{frac--a3} configuration (lower row). For each percentile, wide light bars correspond to low-sensitivity shots (0.90 $ \le \text{sensitivity} \leq$ 0.96) and narrow dark bars to high-sensitivity shots ($\text{sensitivity} >$ 0.96). Red dots indicate the mean bias computed over all shots with $\text{sensitivity} \geq$ 0.90, connected by a line to highlight the overall bias profile. The annotation in each panel reports the mean absolute bias averaged over the full RH1--RH99 range as a global summary measure. All RH metrics are referenced to the ALS-derived ground elevation.}
\label{fig:bias_count_a3}
\end{figure}

\clearpage
\begin{figure}[htbp]
\centering
\includegraphics[width=\textwidth]{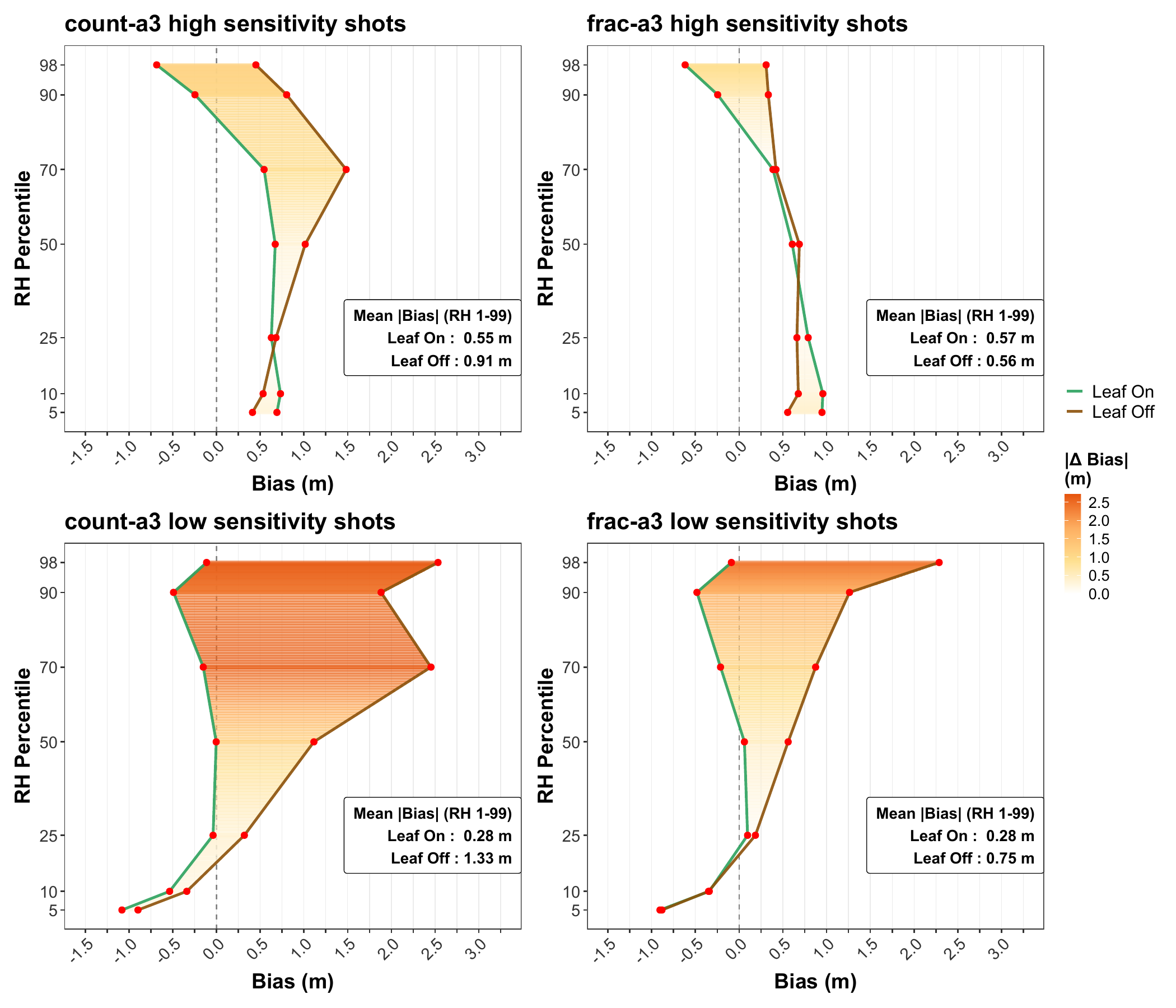}%
\caption{Vertical bias profiles (bias\,=\,mean(simulated RH $-$ aligned GEDI RH)) for leaf-on (green) and leaf-off (brown) conditions under the \textit{count--a3} baseline (left) and the \textit{frac--a3} configuration (right), restricted to broadleaf and mixed forest types. The shaded area and its colour gradient encode the absolute difference in bias between the two seasons ($|\Delta\text{Bias}|$); darker red indicates larger seasonal divergence. Lines correspond to the bias at the displayed percentiles. Overlaid annotation reports the mean absolute bias averaged over the full RH1--RH99 range, providing an overall summary measure. RH metrics are referenced to the ALS-derived ground elevation.}
\label{fig:bias_phenology_count_a3}
\end{figure}

\end{document}